\documentclass[article]{JHEP3}
\usepackage{amsmath,amssymb}
\usepackage{epsfig,multicol}

\newcommand{\be}{\begin{equation}}
\newcommand{\ee}{\end{equation}}
\newcommand{\bea}{\begin{eqnarray}}
\newcommand{\eea}{\end{eqnarray}}
\newcommand{\bean}{\begin{eqnarray*}}
\newcommand{\eean}{\end{eqnarray*}}

\def\be{\begin{equation}}
\def\ee{\end{equation}}
\def\ba{\begin{array}}
\def\ea{\end{array}}
\def\bea{\begin{eqnarray}}
\def\eea{\end{eqnarray}}
\def\beas{\begin{eqnarray*}}
\def\eeas{\end{eqnarray*}}

\def\a{\alpha}
\def\b{\beta}

\def\m{\mu}
\def\n{\nu}
\def\la{\lambda}

\def\g{\gamma}

\def\ol#1{{\overline{#1}}}

\def\delsl{\not\!\partial}

\def\beq{\begin{equation}}

\def\eeq{\end{equation}}

\relax

\preprint{ITFA--2002--36\\ LPTENS--02--48\\ HUTP--02/A035\\ \texttt{hep-th/0209164}}

\title{$D$--Brane Dynamics in Constant Ramond--Ramond Potentials,
$S$--Duality and Noncommutative Geometry}

\author{Lorenzo Cornalba$^{1}$, Miguel S. Costa$^{2}$ and Ricardo 
Schiappa$^{3}$
\\
$^{1}$Instituut voor Theoretische Fysica, Universiteit van Amsterdam,\\
Valckenierstraat 65, 1018 XE Amsterdam, The Netherlands\\
\\
$^{2}$Laboratoire de Physique Th\'eorique, \'Ecole Normale
Sup\'erieure,\\ 
F--75231 Paris Cedex 05, France\\
\\
$^{3}$Department of Physics, Harvard University,\\
Cambridge, MA 02138, USA\\
\\
\email{lcornalb@science.uva.nl}, \, \email{miguel@lpt.ens.fr}, \, 
\email{ricardo@lorentz.harvard.edu}
}

\abstract{
We study the physics of $D$--branes in the presence of constant 
Ramond--Ramond potentials. In the string field theory context, we first 
develop a general formalism to analyze open strings in gauge trivial 
closed string backgrounds, and then apply it both to the RNS string and 
within Berkovits' covariant formalism, where the results have the most 
natural interpretation. The most remarkable finding is that, in the 
presence of a $Dp$--brane, both a constant parallel NSNS $B$--field and
R--R $C^{\left(p-1\right)}$--field \textit{do not solve the open/closed
equations of motion, and induce the same non--vanishing open string 
tadpole}. After solving the open string equations in the presence of this 
tadpole, and after gauging away the closed string fields, one is left with 
a $U\left(1\right)$ field strength on the brane given by 
$F=\frac{1}{2}\left( B - \star C^{\left( p-1\right)}\right)$, where $\star$ 
is Hodge duality along the brane world--volume. One 
observes that this result differs from the usually assumed result 
$F=B$. Technically, this is due to the fact that supersymmetric and 
bosonic string world--sheet theories are different. Note, however, 
that the usual $F+B$ combination is \textit{still} the combination which
remains gauge invariant at the $\sigma$--model level. Also, the standard result $F=B$ is, in the $D3$--brane case, \textit{not compatible with $S$--duality}. On the other hand our result, which is derived automatically given the general
formalism, offers a non--trivial check of $S$--duality, \textit{to all orders 
in} $F$, and this leads to an $S$--dual invariant Moyal deformation. In 
an appendix, we solve the source equation describing the open superstring in a 
generic NSNS and RR closed string background, within the super--Poincar\'e covariant formalism.
}


\begin{document}



\vfill

\eject

\section{Introduction and Motivation}

\label{IM}

The dynamics of $D$--branes in a constant Neveu--Schwarz--Neveu--Schwarz
(NSNS) $B$--field has been extensively analyzed in the literature. It leads
to non--trivial physics on the brane which can be described, at low
energies, by noncommutative Yang--Mills theory. In this paper we wish to
analyze the physics of open strings in the presence of \textit{gauge trivial}
Ramond--Ramond (RR) potentials. At first sight one expects the dynamics of
strings to be completely unaltered by the presence of gauge trivial
potentials, since the string does not carry any RR charge. On the other
hand, the following arguments based on $S$--duality suggest that the issue
is more subtle, and that one should expect, in certain cases, physical
effects.

As a first simple example of the kind of questions we wish to address, let
us consider a $D9$--brane in a flat type IIB background. Let us suppose that
we turn on a gauge trivial NSNS $B$--field $B=d\lambda $, which we assume to
be \textit{localized in spacetime}. From the closed string point of view
nothing is changed, since the new background is a gauge transformation of
the standard flat background. On the other hand, the open string equations
of motion are not satisfied in general, unless we turn on a $U(1)$ potential 
$a_{m}$ on the brane which must satisfy 
\begin{equation}
\partial ^{m}f_{mn}=J_{n},  \label{basicEQ}
\end{equation}
where 
\begin{equation*}
J_{n}=-\partial ^{m}B_{mn}=-\square \lambda _{n}+\partial _{n}\left(
\partial \cdot \lambda \right) .
\end{equation*}
Equation (\ref{basicEQ}) is nothing but Maxwell equations coupled to a
conserved current $J_{m}$, and it is therefore natural to assume that the
physically relevant solution is the one obtained using retarded propagators.
The solution in Lorentz gauge just reads 
\begin{equation*}
a_{n}=\frac{1}{\square \pm i\epsilon }J_{n}=-\lambda _{n}+\partial
_{n}\left( \frac{\partial \cdot \lambda }{\square \pm i\epsilon }\right) 
\end{equation*}
and we therefore easily conclude that $f=-B$ and that the brane $U(1)$ field
strength screens completely the perturbation due to $B$. This process can be
understood pictorially following figure \ref{fig}, where we show on the left
the original $B$--field and in the center the induced current which
behaves---as shown on the right---as a pair of capacitor plates creating a
field $f$ equal and opposite to the original $B$--field. Note that, since
the low--energy effective action is written in terms of the field $f+B$, the
general solution to (\ref{basicEQ}) is clearly $f=-B+\delta f$, where $%
\delta f$ satisfies the free Maxwell equations. The basic point of this
first example is that, \textit{if we consider the D--branes as reacting to a
pre--existing closed string background, and if they do so following a causal
retarded propagation of the world--volume fields, then we single out the
specific solution with} $\delta f=0$.

A second, more physical, example is given by the dynamics of a $D3$--brane
positioned at $x^{4}=\cdots =x^{9}=0$, in the presence of a $B$--field shock
wave, given by 
\begin{equation*}
B=\phi \left( x^{+}\right) dx^{2}\wedge dx^{3}\,,
\end{equation*}
where $x^{\pm }=x^{0}\pm x^{9}$. We choose the function $\phi \left(
x^{+}\right) $ to interpolate from $\phi \left( -\infty \right) =0$ to $\phi
\left( +\infty \right) =\phi $. The NSNS field strength 
\begin{equation*}
H=\phi \,^{\prime }\,dx^{+}\wedge dx^{2}\wedge dx^{3}
\end{equation*}
satisfies $d\star H=0$ for any choice of the function $\phi \left(
x^{+}\right) $ so that, by turning on the $B$--field infinitesimally slowly
(adiabatically), we can solve the closed equations of motion to arbitrary
accuracy without deviating from the flat metric. It is easy to see that the
current $J_{a}$ vanishes in this particular example\footnote{%
Throughout the paper we consider type IIB theory in flat space, together
with a $Dp$--brane stretched in the directions $x^{0},x^{1},\ldots ,x^{p}$.
We use indices $m,n,\cdots =0,\ldots ,9$, for spacetime, $a,b,\cdots
=0,\ldots ,p$, for the brane worldvolume and $i,j,\cdots =p+1,\ldots ,9$,
for the transverse directions.} and therefore, following again the
prescription described in the previous example, we have that $f_{ab}=0$. The
full solution then adiabatically interpolates between a vanishing $B$--field
and a constant field $B=\phi \,dx^{2}\wedge dx^{3}$ on the brane, which
leads to the usual noncommutative behavior of the brane gauge theory.

\EPSFIGURE[t]{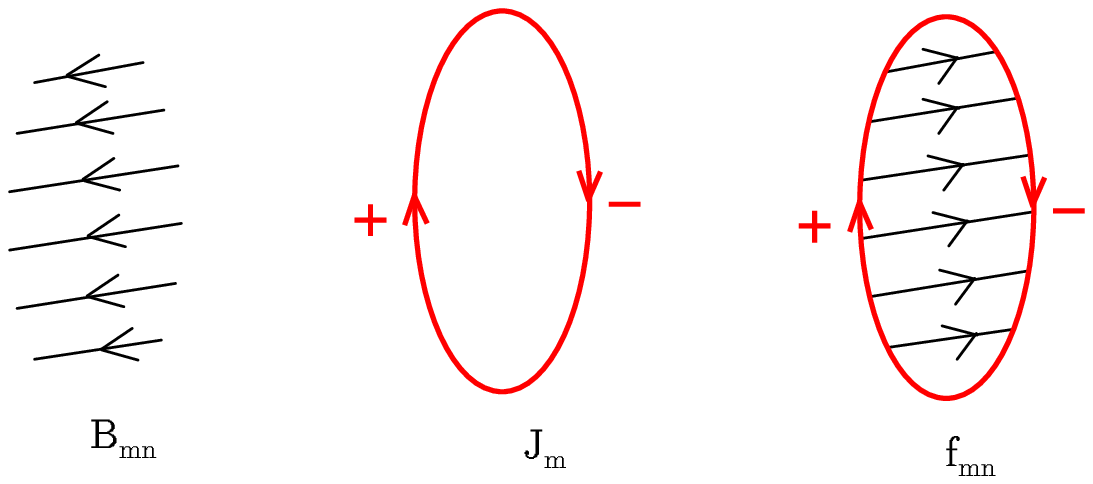}{Backreaction of the open string field $f_{mn}$ to 
the presence of a current $J_m$ generated by a localized gauge trivial 
NSNS $B$--field.\label{fig}}

Let us consider also the $S$--dual description of the above process. This
involves looking at the RR shock wave 
\begin{equation*}
C^{\left( 2\right) }=-\phi \left( x^{+}\right) dx^{2}\wedge dx^{3}\,.
\end{equation*}
In this case there is an induced current $J_{m}$, due to the Wess--Zumino
coupling $-\int_{D3}C^{\left( 2\right) }\wedge f$, which is non vanishing
and is 
\begin{equation*}
J=\phi ^{\prime }(x^{0})\,dx^{1}\,.
\end{equation*}
The Maxwell problem (\ref{basicEQ}) is simply solved, \textit{again with
retarded propagators}, by 
\begin{equation*}
f=\phi (x^{0})\,dx^{0}\wedge dx^{1}\,,
\end{equation*}
so that one now sees that the noncommutativity in the future is due to the
reaction of the open string $U(1)$ gauge potential to the current induced by
the non vanishing RR field.

In these two examples, we have considered the process of turning on some
closed string fields (always satisfying the equations of motion for the
closed string) which are \textit{time dependent and which vanish in the far
past}. These fields induce a current $J$ which acts as a source for the open
string fields, and we analyzed the \textit{unique} reaction of the brane to
this current by solving the corresponding Maxwell equation (\ref{basicEQ})
using the \textit{physical requirement of causality, and therefore using
retarded propagators}. The example of the shock wave is particularly
illuminating: one observes that, in order to obtain the standard
noncommutative behavior to the future of the $S$--dual RR wave, which is
expected by $S$--duality, we must incorporate the solution to equation (\ref
{basicEQ}) due to the current $J_{m}$. We should point out at this stage
that the only principle we follow is a basic prescription on how to solve
the open string equations \textit{uniquely, given a current }$J$. We then
notice that if we turn on two closed string field configurations, which are $%
S$--dual, then the final backgrounds including the backreaction of the brane
will also be $S$--dual, as one expects.

The situation is not as straightforward when the closed string deformation
does not vanish in the far past, and in particular is \textit{constant in
spacetime}. We shall thus mostly focus, from now on, on gauge trivial closed
string deformations. We have seen with the first example that, for gauge
trivial deformations which are localized in spacetime, nothing happens after
one takes into account the open strings backreaction. On the other hand, as
was just mentioned, we are interested in closed string deformations which
are constant in spacetime. In the presence of a constant current, one does
not have a simple physical principle (like retarded propagation of
world--volume fields) to determine the backreaction of the open string,
aside from the only requirement that:

\begin{description}
\item  (A) \textit{Whenever the induced current on the brane is the same,
then the open string reaction should also be identical.}
\end{description}

\noindent We have seen that, in the shock wave example, the final result
compatible with (A) was also compatible with $S$--duality, in the sense that:

\begin{description}
\item  (B) \textit{Given two deformations of the closed string background
which are }$S$\textit{--dual then, after inserting a }$D$\textit{--brane,
the total backgrounds---i.e., the ones which include the backreaction of the
open strings---are also }$S$\textit{--dual.}
\end{description}

\noindent On the other hand, just on the basis of Born--Infeld theory, it
seems impossible to reconcile requirements (A) and (B) whenever the closed
string perturbation is independent of time, as the following example shows.
Consider a $D3$--brane in flat space stretched in the directions $0,1,2,3$,
and let us turn on an NSNS field $B_{23}$. The current $J_{a}$ vanishes and
we then use (A) to insist that the backreaction of the open strings should
also vanish. The full system will then exhibit noncommutative behavior in
the scattering processes of open string states, in the form of the
well--known Moyal phase factors. The $S$--dual closed string background is,
on the other hand, a constant RR $2$--form field $C_{01}$, which again
induces no current. This time, however, the dynamics on the brane is
completely unaltered since the background $U(1)$ field strength vanishes and
since the Wess--Zumino action is a total derivative for gauge trivial RR
fields, and therefore does not contribute to perturbative scattering
amplitudes, unlike the previous situation.

In this paper we shall show that, if we consider the full string fields
(open and closed) as opposed to only the low energy fields (meaning the
massless fields in the low energy effective action with massive modes
integrated out), then the situation is quite different and is in fact 
\textit{compatible with $S$--duality, in the sense of} (B). In particular we
shall show, both in the RNS and in the pure spinor covariant formulation of
Berkovits, that constant NSNS and RR fields \textit{do induce a non
vanishing current, which is the same for both the $B$ and the $C$ field
backgrounds}. Following (A) we must then only consider a single reaction
from the open strings. We will then show, quite non--trivially, that the
resulting total deformation of the background will be \textit{compatible
with $S$--duality independently of the initial configurations of closed and
open strings}. More precisely, we will consider a $Dp$--brane stretched in
directions $x^{0},\cdots ,x^{p}$ with an initial background $U\left(
1\right) $ flux $F_{ab}$, and we will consider turning on constant NSNS and
RR fields $B$ and $C^{\left( q\right) }$. After computing the backreaction
of the open strings following (A), we consider the final configuration
which, quite naturally, corresponds to the original brane with a new gauge
field 
\begin{equation*}
F_{ab}+\delta F_{ab}\left( F,B,C^{\left( q\right) }\right) ,
\end{equation*}
and possibly also with a rotated world--volume. We will then show that, in
the case of a $D3$--brane, the result $\delta F_{ab}$ is compatible with $S$%
--duality, independently of $F$. To leading order, for fields $B$ and $%
C^{\left( 2\right) }$ parallel to the brane, one has ($\star$ in here is
Hodge duality on the brane) 
\begin{equation*}
\delta F=\frac{1}{2}\left( B-\star C^{\left( 2\right) }\right) + \mathcal{O}%
\left(F\right) ,
\end{equation*}
which is compatible with $B\rightarrow C^{\left( 2\right) },C^{\left(
2\right) }\rightarrow -B$ and $\delta F\rightarrow \star \delta F$. The
higher order terms in $\mathcal{O}\left( F\right)$ are computed explicitly
and provide a strong check on the full construction.

This expression corresponds to the linear terms in an NSNS and RR
noncommutative parameter $\Theta $. In order to compute the full nonlinear
result one has to solve a complicated differential equation. Let us also
comment on the factor of $\frac{1}{2}$ (which makes our result differ from
the lore that one should have $F=B$). This factor arises due the fact that
the world--sheet theory for the superstring (be it RNS or pure spinor) is
quite different from the world--sheet theory for the bosonic string where
there is no RR sector, the current $J$ vanishes, and one has $F=B$.
Moreover, we should point out that the canonically normalized $F+B$
combination is \textit{still} the $\sigma $--model gauge invariant
expression to consider, as we show later in the paper.

This paper is organized as follows. We begin in section 2 with a brief
summary of our results, for the reader who wishes to skip the technical
details at first reading. In section 3 we describe the string field
theoretic setting which allows us to describe open string physics in gauge
trivial closed string backgrounds. This setting will be the basis for the
calculations we shall perform in the following sections. We begin the
calculations within the familiar RNS language \cite{FMS, CFQS, KLLSW, KLS,
Klebanov-Thorlacius, GHKM, Garousi-Myers, Hashimoto-Klebanov, BIANCHI,
VFPSLR, BVFLPRS, Polyakov}, in section 4. We shall focus our study on the
case of a $D9$--brane in a constant RR $C$--field and observe that the
boundary deformation, arising from the closed string background in the open
string disk diagrams, precisely corresponds to that of an effective $\delta
F $, as previously described. In order to analyze the most general picture,
we switch to the pure spinor covariant formalism \cite{Siegel, Berkovits-1,
Berkovits-Vallilo, Berkovits-2, Berkovits-3, Berkovits-Chandia-1,
Berkovits-4, Berkovits-Chandia-2, Berkovits-Howe, Berkovits-5,
Berkovits-Chandia-3, Berkovits-Pershin, Berkovits-6, GPN1, GPN2, GPN3, GPN4}
in section 5. The power of this formalism allows us to fully describe $D$%
--brane physics in constant NSNS and RR potentials. The analysis of both
sections 4 and 5 is done for background field strength $F_{ab}=0$ on the
brane. In section 6 we consider the general case of finite $F_{ab}$, and set
up the differential equation $\delta F_{ab}\,$ to determine the full
nonlinear corrections to the background $F_{ab}$ and therefore to the open
string parameters $G$ and $\Theta $ \cite{ACNY, CDS, Cheung-Krogh, Chu-Ho,
Schomerus, Cornalba-Schiappa-1, Seiberg-Witten, Cornalba-2, Chu-Zamora}.
This last result then gives us the tools to perform the non--trivial check
of $S$--duality in the $D3$--brane case, which we have discussed at length
\footnote{Previous work concerning $S$--duality in the context of noncommutative
geometry can be found in \cite{GRS,GMMS,NOS}.}. We end in section 7 with
some conclusions and open problems for future research. In appendix B
it is shown how to solve a source equation describing the motion of an open
superstring in a generic NSNS and RR closed string background.

\section{Summary of Results}

Let us summarize the main results we obtain in this paper. All formulae
apply to type IIB theory, for infinitesimal $B$ and $C$--fields. The
extension to type IIA is trivial.

We first consider a $Dp$--brane with \textit{no background} $U\left(
1\right) $ \textit{field strength} $F=0$, immersed in a parallel $B$ field
and in a general but constant RR potential 
\begin{equation*}
C=C_{\parallel }^{(q)}\wedge C_{\perp }^{(k)}\ ,
\end{equation*}
\noindent where $C_{\parallel }^{(q)}$ and $C_{\perp }^{(k)}$ are,
respectively, a $q$ and a $k$--form parallel and transverse to the brane.
Let also $\star $ be Hodge duality \textit{on the brane world--volume}. The
only non--trivial open string physics coming from the RR sector arises for
the cases $(q,k)=(p-1,0)$ and $(q,k)=(p,1)$. In the first case, the
two--form $\star C_{\parallel }^{(p-1)}$ is parallel to the brane and can be
gauged away, together with the $B$--field, to a constant field strength 
\begin{equation*}
\delta F=\frac{1}{2}\left( B-\star C^{\left( p-1\right) }\right) \ ,
\end{equation*}
on the brane world--volume. Note that the normalization of the above fields
is canonical, and that the non--standard factor of $1/2$ differs from the
usually assumed result $\delta F=B$. This leads to the usual Moyal
deformation \textit{also} for RR fields and is compatible with $S$--duality,
as discussed in the previous section. The other non--trivial case $%
(q,k)=(p,1)$ corresponds to a rotation of the brane world--volume. More
precisely, if we define the two--form 
\begin{equation*}
\Omega =\star C_{\parallel }^{(p)}\wedge C_{\perp }^{(1)}\ ,
\end{equation*}
\noindent and if $x^{a}$ are the world--volume coordinates and $z_{i}$ the
transverse scalars, then the displacement is 
\begin{equation*}
z_{i}=\frac{1}{2}\,x^{a}\,\Omega _{ai}\,.
\end{equation*}

We also consider $Dp$--branes in the presence of a finite background $%
U\left( 1\right) $ field strength $F$. In the bulk of the paper we describe
the general result, but here let us discuss the most relevant case, for $p=3$
with \textit{parallel} background fields $B$ and $C^{\left( 0\right) }$, $%
C^{\left( 2\right) }$ and $C^{\left( 4\right) }$. We discovered that the
total induced variation $\delta F$ of the field strength $F$ is given by 
\begin{equation*}
\delta F\left( F,B,C^{\left( 0\right) },C^{\left( 2\right) },C^{\left(
4\right) }\right) =\frac{1}{2}\left( B-\star C^{(2)}\right) -\frac{1}{2}%
F\wedge \star C^{\left( 4\right) }+\frac{1}{2}\star F\wedge C^{\left(
0\right) }+\cdots \ ,
\end{equation*}
where the dots represent terms of higher order in $F$ which can be computed
explicitly. The most remarkable feature of the above result is its
compatibility with $S$--duality, in the sense described qualitatively in the
Introduction and to be made precise in section $6.2$. In fact, using the
explicit formulae in this paper, we have checked $S$--duality to \textit{all
orders in} $F$! This leads to an $S$--dual invariant prescription for a
Moyal noncommutative deformation.

\section{Basic String Field Theory Setting}

\label{BSFTS}

In order to understand the issue more clearly, we first need to discuss,
from a general and qualitative point of view, what we mean by the motion of
an open string in a closed string background, and in particular in a gauge
trivial closed string background.

In general, strings at weak coupling are described by an open--closed string
field theory action \cite{Zwiebach-1, Gaberdiel-Zwiebach-1, Zwiebach-2,
Gaberdiel-Zwiebach-2} which is written in terms of a closed and an open
string field, $\Psi $ and $\Phi $ (in bosonic string theory, respectively of
ghost number $2$ and $1$), and which is of the general form 
\begin{equation*}
\frac{1}{g^{2}}S_{c}+\frac{1}{g}S_{o}+\frac{1}{g}S_{oc}\ ,
\end{equation*}
\noindent where $S_{c}$ represents the closed string field theory action,
coming from interactions with sphere diagrams, and where $S_{o}$ and $S_{oc}$
correspond to the open string field action and to the interaction terms
between open and closed strings, and are related to disk diagrams. This
explains the different powers of the string coupling in front of the various
terms. Focusing on the linearized classical equations of motion in the free
$g\rightarrow 0$ limit, we have in general the equations 
\begin{eqnarray}
Q\Psi &=&0\ ,  \label{eq200} \\
Q\Phi &=&-J\ ,  \label{eq300}
\end{eqnarray}
where $Q$ is the (either open or closed) BRST\ operator, and \noindent where
we have defined 
\begin{equation*}
J\equiv \pi \left( \Psi \right) .
\end{equation*}
\noindent The map $\pi $ relates closed to open string states. One can think
of this map as a projection map, projecting the closed string vertex
operators in the world--sheet bulk to the world--sheet boundary
(implementing appropriate boundary conditions between left and right
movers). Its detailed properties depend on the specific string field theory
in question, but $\pi $ generically will preserve ghost number and will have
the property that 
\begin{equation*}
Q\pi \left( \Xi \right) =\pi \left( Q\Xi \right) .
\end{equation*}
\noindent It is clear that equations (\ref{eq200}) and (\ref{eq300}) are
asymmetric and one can, in particular, choose $\Psi $ to satisfy the closed
string equations of motion completely forgetting about the open string
sector (this is true also in the interacting theory). This asymmetry is
directly related to the different powers of $g$ corresponding to closed and
open string interactions, and it is the very reason why one can talk about
open strings in a closed string background but not vice--versa. Given a
solution, $\Psi $, in order to find a consistent vacuum one must solve
equation (\ref{eq300}) for $\Phi $. We now recall that, for the massless
bosonic sector, the on--shell equation $Q\Phi =0$ corresponds to the \textit{%
free} Maxwell equations of motion. This implies that we should consider $J$
as a source term for the Maxwell equations, or as a \textit{current}. Then,
current conservation is nothing but the statement that $QJ=\pi \left( Q\Psi
\right) =0$. These facts are the generalization, using the full string
field, of the setup described in the introduction, and we shall see, in the
following sections, concrete examples both in the RNS and in the pure spinor
formalism.

Let us now consider the gauge invariance of the string field theory
equations of motion, focusing on the linear part of the gauge
transformations (since we are looking only at the linear equations of motion
in this discussion). We clearly have the usual open string gauge
transformations $\Phi \rightarrow \Phi + Q\kappa$. We will be, on the other
hand, more interested in the closed string gauge transformations which read,
in the presence of the open string sector, 
\begin{equation}
\Psi \rightarrow \Psi -Q\eta\ ,\ \ \ \ \ \ \ \ \ \ \ \ \ \ \ \ \ \ \ \ \ \ \
\ \ \ \ \ \ \Phi \rightarrow \Phi +\pi \left( \eta \right)\ .  \label{eq600}
\end{equation}
\noindent Clearly any two vacua related by the above gauge transformations
will be completely equivalent and will yield the same physical results.

We may now discuss more clearly what we mean by open strings propagating in
a gauge trivial closed background. Consider a closed string background given
by 
\begin{equation*}
\Psi =Q\eta \ .
\end{equation*}
\noindent As we pointed out, one can choose this background \textit{before}
talking about open strings. As a second step we need to solve equation (\ref
{eq300}) for the open string field. Denoting the solution with $\Phi _{r}$,
we need to solve 
\begin{equation}
Q\Phi _{r}=-J=-Q\Phi _{c}\ ,  \label{eq500}
\end{equation}
\noindent where we have defined the open string state 
\begin{equation}
\Phi _{c}\equiv \pi \left( \eta \right) .  \label{eq800}
\end{equation}
\noindent Clearly one solution of (\ref{eq500}) is given by $\Phi =-\Phi
_{c} $. Recalling the first example in the introduction, this solution will
be the natural one whenever $\Psi $ is localized in spacetime. This case is,
on the other hand, quite uninteresting since the vacuum $\Psi =Q\eta $, $%
\Phi =-\pi \left( \eta \right) $ is a gauge transform of the $\Psi =\Phi =0$
vacuum. For more general configurations, as we have already discussed at
length in the introduction, we must have a general principle on how to solve
(\ref{eq500}), which must be compatible at least with requirement (A) in the
introduction. At any rate, we will later discuss more generally how to solve
equation (\ref{eq500}), but for now we just denote with $\Phi _{r}$ the
solution, \textit{which we assume unique given} $J$. We have the new, in
general \textit{physically different} vacuum $\Psi =Q\eta $, $\Phi =\Phi
_{r} $. We can then apply the gauge transformation (\ref{eq600}) to bring
the solution to the form 
\begin{eqnarray*}
\Psi &=&0\,, \\
\Phi &=&\Phi _{c}+\Phi _{r}\,.
\end{eqnarray*}
\noindent In this way, we can associate to \textit{any} gauge trivial closed
string state $\Psi =Q\eta $ a corresponding \textit{on--shell} deformation
of the \textit{open string} $\Phi _{c}+\Phi _{r}$ by using equations (\ref
{eq800}) and (\ref{eq500}). We therefore have a map which sends gauge
trivial closed string deformations into on--shell physical open string
deformations. We want to show that, whenever $\Psi $ and $\widetilde{\Psi }$
are $S$--dual closed gauge--trivial deformations, then our procedure will
yield $S$--dual open string deformations $\Phi _{c}+\Phi _{r}$ and $%
\widetilde{\Phi }_{c}+\widetilde{\Phi }_{r}$, under the requirement (A) that 
\textit{equal currents $J$ go into equal solutions $\Phi _{r}$ of (\ref
{eq500})}. We will show that this is possible for the case of constant NSNS
and RR gauge field backgrounds. A word on the notation. We have divided the
open string field $\Phi $ in a part coming from the closed string and a part
coming from the reaction of the open string. This explains the choice of
subscripts $\Phi _{c}$ and $\Phi _{r}$.

To illustrate this procedure in a familiar setting, let us consider the bosonic string and show how to obtain the well--known (infinitesimal) deformation of $F=B$. This will make clear, in a simplified case, how the above reasoning works. Given the constant $B$--field bosonic closed string background, described at zero momentum by
$$
\Psi = \frac12\, c \overline{c} \left[ B_{mn} \partial X^{m} \overline{\partial} X^{n} \right],
$$
\noindent
it is simple to compute that the corresponding current will vanish, $J=\pi(\Psi)=0$. Indeed, the $\pi$ map brings the closed string vertex operator to the boundary of the world--sheet, where $c=\overline{c}$. Following the procedure outlined above one now computes, in a unique way, the open string reaction via $Q \Phi_{r} = - J = 0$, which yields $\Phi_{r} = 0$. The open string deformation will thus be given by $\Phi = \Phi_{c} + \Phi_{r} = \Phi_{c}$, where one still needs to find $\Phi_{c} = \pi (\eta)$. First observe that with
$$
\eta = \frac14\, B_{mn} \left[ X^{m}\, \overline{c} \overline{\partial} X^{n} - c \partial X^{m}\, X^{n} \right] 
$$
\noindent
it follows that $\Psi=Q\eta$. Then, it is simple to compute the open string deformation as
$$
\Phi = \pi(\eta) = \frac12\, B_{mn}\, X^{m} \partial X^{n}.
$$
\noindent
This is the expected final result of $F=B$ open string deformation, for the bosonic string. In the superstring case one will find that the current $J$ is actually non--vanishing (though equal in the two cases of NSNS and RR deformations), and the whole story will be very different from this simple bosonic case. In fact, $\Phi_{c}$ alone will no longer be a solution of the open string equations of motion, and one will have to endure a much harder analysis in order to add the $\Phi_{r}$ contribution.

We now have the general framework to address our main problem at hand. We need to
understand better, in specific cases, the map $\pi$ and, more importantly,
how to solve equation (\ref{eq500}) or, more generally, equation (\ref{eq300}). 
We shall first start by performing such an analysis in the familiar
setting of the RNS formalism, and will only later proceed to use the more
powerful pure spinor covariant formalism, where the whole picture will
become much clearer.

\section{Analysis in the RNS Formalism}\label{ARNSF}

In the RNS formulation of the superstring \cite{FMS, CFQS, KLLSW, 
KLS} the matter fields organize into the action
$$
\frac{1}{2\pi} \int d^{2}z\ \left( \frac{2}{\alpha'}
\partial X^{\mu} \bar{\partial} X_{\mu} + \psi^{\mu} \bar{\partial}
\psi_{\mu} + \overline{\psi}^{\mu} \partial \overline{\psi}_{\mu} \right),
$$
\noindent
whereas the ghost fields are the usual $bc$ and $\beta\gamma$ systems. It is 
standard to fermionize the $\beta\gamma$ superconformal ghosts as 
$\beta = \partial \xi e^{-\phi}$ and $\gamma = \eta e^{\phi}$. Also, we shall 
work in $\alpha' = 2$ units. The BRST operator $Q=Q_L + Q_R$
naturally decomposes into three pieces \cite{FMS},
\begin{equation} \label{QBRSTRNS}
Q_{L} = Q_{0} + Q_{1} + Q_{2}\ ,
\end{equation}
\noindent
labeled by their spinor ghost charges
\begin{eqnarray*}
Q_{0} &=& \oint \frac{dz}{2\pi i}\ \left[ c \left( T_{X}  + T_{\psi}
+ T_{\eta\xi} + T_{\phi} \right) + c \partial c b
\right]\ , \\
Q_{1} &=& \frac{1}{2} \oint \frac{dz}{2\pi i}\ e^{\phi} \eta
\psi^{\mu}  \partial X_{\mu}\ , \\
Q_{2} &=& \frac{1}{4} \oint \frac{dz}{2\pi i}\ b \eta \partial
\eta e^{2\phi}\ .
\end{eqnarray*}
\noindent
The BRST operator has total ghost number $1$ and total picture number $0$.

\subsection{Vertex Operator $\Psi$ for the $B$ and $C$--Fields in Superstring 
Theory}

In this section we analyze the vertex operators $\Psi_B$ and $\Psi_C$ 
for a constant $B$--field and a constant $C$--field.

We recall first that scattering amplitudes on the disk are saturated 
with the insertion of three ghost fields $c,\overline{c}$ and with a total
picture number of $(-2)$ \cite{KLS}. In particular, for closed string
vertex operators, picture number is counted by summing the left and
right sector picture numbers $p$, $\ol{p}$. We shall subsequently be 
interested in world--sheet bulk vertex operators $\Psi$ such that 
$p+\overline{p} = -2$. Therefore, if one considers the disk interaction of 
$n$ gluons and $2m$ gluinos, in the presence of the closed vertex $\Psi$, the 
total picture number for the gluons and the gluinos must vanish and the 
correlator will look schematically like
$$
\langle\; \underbrace{ {\mathcal{V}}_{(0)}\; \cdots\; 
{\mathcal{V}}_{(0)}}_{n}\; \underbrace{ {\mathcal{V}}_{(+\frac{1}{2})}\; 
{\mathcal{V}}_{(-\frac{1}{2})}\; \cdots\; {\mathcal{V}}_{(+\frac{1}{2})}\; 
{\mathcal{V}}_{(-\frac{1}{2})} }_{2m}\; \Psi \; \rangle\ .
$$

The $B$--field vertex operator in the $(-1,-1)$ picture is standard and is 
given, at zero momentum, by the expression
\begin{equation} \label{VB}
\Psi^B = - c \overline{c} \left[ B_{mn} \psi^{m} 
\overline{\psi}^{n} e^{-\phi} e^{-\overline{\phi}}
\right].
\end{equation}
\noindent
Consider now the $C$--field bulk vertex operator. In the standard
$(-\frac{1}{2},-\frac{1}{2})$ symmetric picture this vertex operator will 
contain the RR field strength $F_{q+1}=dC_{q}$, and not the RR gauge 
potential $C_{q}$, as is well known, for example, from many $D$--brane 
scattering calculations \cite{Klebanov-Thorlacius, GHKM, Garousi-Myers, 
Hashimoto-Klebanov}. In order to obtain a vertex operator which depends on 
the RR gauge potential one needs to change to an asymmetric picture 
\cite{BIANCHI, VFPSLR, BVFLPRS}. Due to the nature of our calculations, this 
asymmetric picture vertex operator is precisely the one we are interested in. 
It has, as we shall see, total picture number $(-2)$ and it is therefore the
natural companion of the operator $\Psi^B$ for constant $B$--field. We will 
present explicit formulae for type IIB, the extension to IIA being trivial.

Let us first introduce some notation for the RR potentials. For spinor 
conventions, please refer to appendix A. We let $C_{m_1\cdots m_q}$, with $q$ 
even, be the RR $q$--form potential, and we define as usual the bi--spinor
$$
C_{\alpha }{}^{\beta } = \sum_{q\mathrm{\ even}}\frac{1}{q!}C_{m_{1}\cdots
m_{q}}\left( \gamma ^{m_{1}\cdots m_{q}}\right) _{\alpha }{}^{\beta }\ ,
$$
\noindent
and similarly for the dual version $C^{\alpha}{}_\beta$. Also, we will find 
convenient to use, together with $C^{\alpha}{}_\beta$, its transpose 
$\ol{C}_\alpha{}^\beta = C^{\beta}{}_\alpha$ given by
$$
\ol{C}_{\alpha }{}^{\beta } = \sum_{q\mathrm{\ even}}\frac{(-)^{\frac{q}{2}}}{q!}C_{m_{1}\cdots
m_{q}}\left( \gamma ^{m_{1}\cdots m_{q}}\right)_{\alpha }{}^{\beta }\ .
$$
\noindent
The $C$--field vertex operator $\Psi^C$, at zero momentum, is a sum of two 
pieces at picture number $(-\frac{3}{2},-\frac{1}{2})$ and 
$(-\frac{1}{2},-\frac{3}{2})$ \cite{BIANCHI, VFPSLR, BVFLPRS}
\begin{equation}\label{WIIB}
\Psi^C = \frac{1}{16} c \overline{c} \left[
S^{\alpha}\ {{C}_{\alpha}}^{\beta}
\overline{S}_{\beta} e^{-\frac{3}{2}\phi}
e^{-\frac{1}{2}\overline{\phi}}
-
S_{\alpha}\ {{C}^{\alpha}}_{\beta}
\overline{S}^{\beta} e^{-\frac{1}{2}\phi}
e^{-\frac{3}{2}\overline{\phi}}
\right]\, ,
\end{equation}
\noindent
where $S$ are the spin fields.

In the language of section \ref{BSFTS}, we now have the closed string field 
$\Psi$. It is pure gauge, so we proceed to compute the corresponding 
open string current $J=\pi(\Psi)$ (section \ref{BOPEs}) and the BRST potential
$\eta$ and $\Phi_c=\pi(\eta)$ (section \ref{BRSTP}).

\subsection{Boundary OPE's for $\Psi^B$ and $\Psi^C$: Computation of the 
Current $J$} \label{BOPEs}

In this section we wish to compute the open string current $J=\pi(\Psi)$ 
generated by the closed string field $\Psi$. Therefore we must introduce open 
strings by restricting the CFT to the upper half complex plane ${\mathbb {H}}$ 
and by imposing boundary conditions on $\partial {\mathbb {H}}$. We will then 
define the closed--open projection $\pi(\Psi)$ to be simply the boundary OPE 
of the operator $\Psi$ as it approaches $\partial {\mathbb {H}}$. This 
definition is clearly valid only if the OPE is non--singular, which will be 
true for the applications which follow.

We will discuss in this section the specific case of the $D9$--brane 
in type IIB. The more general cases are discussed in section 
\ref{ACF} using the covariant formalism. The $D9$--brane boundary conditions 
on $\partial {\mathbb{H}}$ are
\begin{eqnarray*}
\overline{\partial} X^m &=& \partial X^m\, ,\ \ \ \ \ \ \ \ \ \ \ \ 
\overline{\psi}^m =  \psi^m\, ,\\
\overline{S}^\alpha &=& S^\alpha\, ,\ \ \ \ \ \ \ \ \ \ \ \ \ \ \ 
\overline{S}_\alpha = S_\alpha\, ,
\end{eqnarray*}
\noindent
together with the boundary conditions for the ghost fields,
$$
\overline{{\mathrm{Ghost}}} = {\mathrm{Ghost}}\, .
$$

Let us start by computing the current $J^B = \pi(\Psi^B)$ related to a 
constant $B$--field background. As we discussed above the definition of 
$\pi$ implies that we must compute the following OPE
$$
J^B = \lim_{z\to\bar{z}} \Psi^B(z,\bar{z})\ .
$$
\noindent
Using the boundary conditions we get that
\begin{eqnarray*}
\lim_{z\to\bar{z}} \Psi^B(z,\bar{z})
&=&
B_{mn} \ \lim_{z\to\bar{z}}\ \left(  c \partial c
(\bar{z}) + \frac{1}{2} \left( z-\bar{z} \right) c \partial^{2} c (\bar{z})
+ {\mathcal{O}} (z-\bar{z})^{2} \right) \cdot \\
&&\cdot
\left( \frac{g^{mn}}{z-\bar{z}} - g^{mn} 
{\partial} \phi (\bar{z}) + \psi^{m}\psi^{n} (\bar{z}) + 
{\mathcal{O}} (z-\bar{z})\ \right) e^{-2\phi( \bar{z} )}\ .
\end{eqnarray*}
\noindent
Besides the standard OPE's \cite{FMS}, we have used the
relation\footnote{We are using the notation $g_{mn}$ for the flat 
\textit{diagonal} spacetime metric throughout this paper.}
$$
\psi^{m} (z) \psi^{n} (w) = \frac{g^{mn}}{z-w} + :
\psi^{m}\psi^{n} : (w)+\cdots\ .
$$
\noindent
The final result for the current $J^B$ reads
\begin{equation} \label{BOPE}
J^B = B_{mn}\
c\partial c \ \psi^{m}\psi^{n} \ e^{-2\phi}\ .
\end{equation}
\noindent
Note that $J^B$ is an open string operator at picture number $(-2)$ 
and ghost number $(+2)$ as expected. 

We now turn to the $C$--field case. The treatment is identical to the 
previous situation, with the exception that the analysis involves the OPE's 
of the spin fields 
$$
S^{\alpha} (z) S_{\beta} (w) \sim (z-w)^{-\frac{5}{4}}\
\delta^{\alpha}{}_{\beta} + (z-w)^{-\frac{1}{4}}\ \frac{1}{2} {\left[ 
\gamma_{mn} \right]^{\alpha}}_{\beta}\ \psi^{m}\psi^{n}
(w) + \cdots\ .
$$
\noindent
We may then compute the limit
\begin{eqnarray*}
\lim_{z\to\bar{z}} \Psi^C (z,\bar{z}) &=& -\frac{1}{16}
{C}_{\alpha}{}^{\beta} \ \lim_{z\to\bar{z}}\ \left( 
c \partial c (\bar{z}) + \frac{1}{2} \left( z-\bar{z} \right) c
\partial^{2} c (\bar{z}) + {\mathcal{O}} (z-\bar{z})^{2} \right) \cdot \\
&&\cdot
\Big( \frac{\delta^{\alpha}{}_{\beta}}{z-\bar{z}} - \frac{3}{2}
\delta^{\alpha}{}_{\beta}\ {\partial} \phi(\bar{z}) + \frac{1}{2}
{\left[ \gamma_{mn} \right]^{\alpha}}_{\beta}\
\psi^{m}\psi^{n} (\bar{z}) + {\mathcal{O}} (z-\bar{z}) \Big) e^{-2\phi 
(\bar{z})}-\cdots\ ,
\end{eqnarray*}
\noindent
where the terms in $\cdots$ are similar and come from the picture number 
$(-\frac{1}{2},-\frac{3}{2})$ part of $\Psi^C$. To get a regular OPE we will 
assume\footnote{This implies that $C^{(0)} = 0$. This case is easily treatable 
in the covariant formalism to follow.} that ${C}_{\alpha}{}^{\alpha}=0$. In 
this case we conclude that
\begin{eqnarray} \label{WIIBOPE}
J^C &=& -\frac{1}{32} 
\left(
{C}_{\alpha}{}^{\beta}
\ {\left[ \gamma_{mn}\right]^{\alpha}}_{\beta}\ -
{C}^{\alpha}{}_{\beta}
\ {\left[ \gamma_{mn}\right]_{\alpha}}^{\beta}\
\right)
\ c\partial c
\ \psi^{m}\psi^{n}  e^{-2\phi }\\
&=&
\frac{1}{32} 
{\rm Tr} \left[
(C+\ol{C})\gamma_{mn}\right]
\ c\partial c
\ \psi^{m}\psi^{n}  e^{-2\phi}\ .
\end{eqnarray}
\noindent
Clearly, the only RR--field that contributes to $J^C$ is the 
$8$--form potential (since the sum $C+\ol{C}$ does not include the $2$--form 
potential). Moreover it is easy to show that $J^C$ has the \textit{same form} 
of $J^B$ with an effective $B$--field given by
$$
B_{\rm eff} = -\frac{1}{64}
{\rm Tr} \left[
(C+\ol{C})\gamma_{mn}\right]dx^m\wedge dx^n
= -\star C^{(8)}.
$$
\noindent
This is the first instance of the results we have discussed in section \ref{IM}. 

Two comments are now in order. Firstly, we are discussing only the 
$D9$--brane case, for clarity of exposition. We will discuss the general case 
in the (simpler) framework of the covariant formalism, which we shall 
describe in section \ref{ACF}. Secondly, we should note that these results are 
only valid to linear order in the background fields. In order to determine the 
full nonlinear RR noncommutativity extra work is required. In fact, using the 
general results in the covariant formalism one can write down a differential 
equation for the noncommutativity parameter, as the background fields $B$ and
$C$ are turned on. When one has only $B$--field the full non--linear result is 
known. We shall discuss the non--linear result with both fields in section 
\ref{NLOSP}.

\subsection{BRST Potentials for Vertex Operators: Computing $\eta$ and 
$\Phi_c$}\label{BRSTP}

In the last section we have studied the open string current $J$. 
Since we are dealing with constant $B$ and $C$--fields, we expect 
that the vertex operators (\ref{VB}) and (\ref{WIIB}) are BRST exact.
In this section we address the question of finding the BRST potential, 
$\eta$, for these operators, satisfying $\Psi = Q\eta$, and then study 
their boundary OPE, $\Phi_c = \pi(\eta)$.

Let us begin with the vertex operator for the $B$--field, (\ref{VB}). 
Counting of conformal dimension, ghost and picture numbers, suggests that one 
should consider the operator of ghost number $(-1)$ (concentrating on the 
left--movers for the moment)
$$
2 c \partial \xi e^{-2\phi} f(X)\, .
$$
\noindent
The only non--trivial commutation is with the $Q_{1}$ component of the 
BRST operator, which will produce
$$
\left[ Q_{1}, 2 c \partial \xi e^{-2\phi} f(X) 
\right] = c \psi^{m} e^{-\phi} \partial_m f(X)\, .
$$
\noindent
If we introduce an one--form $\Lambda$ such that $B=d\Lambda$ with
$\partial\cdot \Lambda=\square \Lambda = 0$ (for example $2\Lambda_n = x^m
B_{mn}$) it is quite easy to show that
$$
\Psi^B = \left[ Q_L + Q_R, \eta^B \right],
$$
\noindent
where
$$
\eta^B  = -2 \Lambda_m(X)\ c \ol{c} 
\left[ \partial \xi e^{-2\phi}
\ \overline{\psi}^{m}  e^{-\overline{\phi}}
+\psi^{m} e^{-\phi}\ \bar{\partial} 
\overline{\xi}  e^{-2\overline{\phi}}
\right].
$$
\noindent
To compute $\Phi_c=\pi(\eta)$ we take the boundary OPE of the BRST potential 
and we get
\begin{equation} \label{boundvb}
\Phi^B_c=\lim_{z\to\bar{z}} \eta^B (z,\bar{z}) = 2 \Lambda_{m} (X)\ 
c\partial c \ \psi^{m}\ \partial \phi \partial \xi\ 
e^{-3\phi}\ .
\end{equation}

Next, we turn our attention to the $C$--field vertex operator (\ref{WIIB}). 
Conformal dimension, and ghost and picture number counting now tell us that 
the natural operator to consider is
$$
2\ c \partial c\ \partial \xi  \partial^{2} \xi \ S^{A}
\ e^{-\frac{7}{2} \phi} \ f(X)\ .
$$
\noindent
This time the relevant non--trivial commutation relation is with the
$Q_{2}$ component of the BRST operator (\ref{QBRSTRNS}), where one finds as 
expected
$$
\left[ Q_{2}, 2\ c \partial c\ \partial \xi  \partial^{2} \xi \ S^{A}
\ e^{-\frac{7}{2} \phi} \ f(X) \right] = c\ S^{A}
\ e^{-\frac{3}{2} \phi}\  f(X)\ . 
$$
\noindent
Thus, if one considers the operator
$$
\eta^C = -\frac{1}{8}
c \partial c\ \partial \xi  
\partial^{2} \xi \ e^{-\frac{7}{2} \phi}\ S^{\alpha}\ C_{\alpha}{}^\beta
\ \overline{c}\ \overline{S}_{\beta} \ e^{-\frac{1}{2} 
\overline{\phi}} + \cdots\ ,
$$
\noindent
(the $\cdots$ are the symmetric term with left and right movers interchanged)
commutation of $Q_L+Q_R$ with this operator then yields the $C$--field vertex 
operator
$$
\Psi^C = \left[ 
Q_L+Q_R , \eta^C \right].
$$
\noindent
As we take the boundary OPE of this BRST potential one finds
\begin{equation} \label{boundwc}
\Phi^C_c=\lim_{z\to\bar{z}} \eta^C (z,\bar{z})
= \frac{1}{32} 
{\rm Tr} \left[
(C+\ol{C})\gamma_{mn}\right]
\ c\partial c \partial^{2} c \ \partial 
\xi \partial^{2} \xi \ \psi^{m}\psi^{n} (z)\ e^{-4\phi}\,  .
\end{equation}
\noindent
Observe that, as expected, both (\ref{boundvb}) and (\ref{boundwc}) live at 
picture $(-2)$.

To conclude the general procedure discussed in section \ref{BSFTS}, we 
must now understand the reaction term $\Phi_r$ of the open strings, and the 
total deformation $\Phi_c+\Phi_r$. In order to do so, we analyze in the next 
section the gluon vertex operator at picture number $(-2)$.

\subsection{The Gluon Vertex Operator at Picture Number $(-2)$}

The boundary operators we have obtained in the previous section are 
$$
a_m (X)\ c\partial c\ \psi^{m}\ \partial \phi \partial \xi
\ e^{-3\phi}\ ,
$$
\noindent
and
$$
f_{mn}(X)\ c\partial c \partial^{2} c\ \partial 
\xi \partial^{2} \xi\ \psi^{m}\psi^{n}\ e^{-4\phi}\ .
$$
\noindent
Recall that the ghost number assignment of $e^{q \phi}$ is $q$ 
\cite{FMS}, so that both these operators are at ghost number $(-1)$ and 
picture number $(-2)$. The question we face is whether picture raising 
will bring a linear combination of these operators to the canonical gluon 
vertex operator at picture number $(-1)$
\begin{equation} \label{VA}
\Phi_{(-1)} =2 a_{m}(X)\ c \psi^{m}\ e^{-\phi}\ .
\end{equation}
\noindent
Let us consider the following ans\"atz for the gluon vertex operator 
at picture number $(-2)$,
\begin{eqnarray*}
\Phi_{(-2)}  &=& k_{1}\ a_m (X)\ c\partial c\ \psi^{m}\ \partial
\phi \partial \xi\ e^{-3\phi}\\
&&
+ k_{2}\ f_{mn}(X)\ c\partial c \partial^{2} c\ \partial 
\xi \partial^{2} \xi\ \psi^{m}\psi^{n}\ e^{-4\phi}\ ,
\end{eqnarray*}
\noindent
where $k_{1}$ and $k_{2}$ are two constants to be determined and, at 
the moment, we are assuming no relation between $a_{m}$ and 
$f_{mn}$. We need to show that the picture raising operation will take 
$\Phi_{(-2)}$ to $\Phi_{(-1)}$ or, more specifically, that
$$
\Phi_{(-1)} = \left[ Q, 2 \xi 
\Phi_{(-2)}  \right].
$$
\noindent
As expected, we will have to impose that $f=da$ as well as 
the on--shell relations $\partial\cdot a = 0 = \square a$ for the gluon field.

We shall begin with the piece proportional to $a_{m}$. Assuming  
$\square a = 0$, it is simple to realize that the commutation with the 
$Q_{0}$ component of (\ref{QBRSTRNS}), in the formula above, will 
vanish. To analyze the commutation with $Q_{1}$, one has to 
use the fermion OPE,
$$
\psi^{m} (z) \psi^{n} (w) = \frac{g^{mn}}{z-w} + :
\psi^{m}\psi^{n} : (w)+\cdots\, .
$$
\noindent
The term which arises from the singular part above will be 
proportional to $\partial\cdot a$, which we take to be zero 
in the Lorentz gauge. The term that arises from the normal ordered piece
is non--vanishing and reads
$$
\left[ Q_{1}, 2 \xi\ a_m \ c\partial c\ \psi^{m}\ \partial
\phi \partial \xi\ e^{-3\phi}
\right]
= - \Big( \partial_{m} a_{n} - 
\partial_{n} a_{m} \Big)\ \xi c \partial c  \psi^{m} 
\psi^{n}  e^{-2\phi}\ .
$$
\noindent
At first sight the above result seems strange, as the right hand side is a 
vertex operator off the small Hilbert space. As we shall see in a 
moment, this term will cancel out in the final expression exactly when 
$\Phi_{(-2)}$ is BRST closed. This is simple to understand since the terms 
proportional to $\xi$ arise, schematically, from 
$[ Q, 2 \xi \Phi_{(-2)} ]\simeq 2 \xi  [ Q, \Phi_{(-2)} ] +\cdots $ and vanish 
when $[ Q, \Phi_{(-2)} ]=0$. As to the commutation with the last component of 
(\ref{QBRSTRNS}), $Q_{2}$, it yields the expected expression for the gluon 
vertex operator at the canonical $(-1)$ picture,
$$
\left[ Q_{2}, 2 \xi\ a_m \ c\partial c\ \psi^{m}\ \partial
\phi \partial \xi\ e^{-3\phi} \right]
= \ a_{m}\ c \psi^{m} e^{-\phi}\ .
$$

Let us now turn to the term proportional to $f_{mn}$ in 
$\Phi_{(-2)}$. Again, with $\square f = 0$, one quickly 
shows that the commutation with the $Q_{0}$ component of the BRST 
operator vanishes. Commutation with the $Q_{1}$ component, on the 
other hand, yields a term proportional to the Yang--Mills equation of 
motion $\partial^{m} f_{mn}$, which vanishes for $f=da$ with an on--shell 
gluon. The final commutation with the $Q_{2}$ component of the BRST operator 
results in an operator that is again off the small Hilbert space, namely
$$
\left[ Q_{2}, 2 \xi(z)\ f_{mn}\ c\partial c \partial^{2} c\ \partial 
\xi \partial^{2} \xi\ \psi^{m}\psi^{n}\ e^{-4\phi} \right] = -2\ f_{mn}\ \xi 
c \partial c  \psi^{m} \psi^{n} e^{-2\phi}\ .
$$
\noindent
Because we are looking for the physical gluon vertex operator, at 
picture $(-1)$, we cannot allow for terms which are outside the small 
Hilbert space, so we require that the two terms proportional to $\xi(z)$ 
cancel with each other. This requirement is fulfilled when
$$
k_{1} + 2 k_{2} = 0
$$
\noindent
and $f=da$. Moreover, under this condition, as we discussed above, the state 
$\Phi_{(-2)}$ is BRST closed. In conclusion, fixing the overall normalization, 
the gluon vertex operator at picture $(-2)$ is simply
\begin{eqnarray*}
\Phi_{(-2)}  &=& \ 2\,a_m (X)\ c\partial c\ \psi^{m}\ \partial
\phi \partial \xi\ e^{-3\phi}\\
&&
-\, f_{mn}(X)\ c\partial c \partial^{2} c\ \partial 
\xi \partial^{2} \xi\ \psi^{m}\psi^{n}\ e^{-4\phi}\ .
\end{eqnarray*}
\noindent 
We shall, moreover, denote with $\Phi_{(-2)}(F)$ the gluon vertex operator 
corresponding to a constant field strength $f_{mn}(X) = F_{mn}$, $2a_m(X) = 
-F_{mn} X^n$. 

To conclude the argument and to prove that, in the presence of a
$C$--field (in particular in the example we are discussing, of a $8$--form 
RR potential), the open strings react and behave as in the presence of an 
effective $B$--field $B_{\rm eff} = - \star C^{(8)}$, we must solve the 
source equation $Q\Phi_r = -J$. We shall later develop a general strategy to 
solve this equation in the covariant formalism, so for the moment we are 
forced to borrow one result from section \ref{ACF}, which is based both
on this general formalism as well as on arguments of $S$--duality. 
More precisely, we will show in section \ref{OSCBF} that, in the presence
of a $B$--field, the total boundary deformation will be
$$
\Phi_c^B + \Phi_r^B = \frac{1}{2}\Phi_{(-2)}(B)\ .
$$
We remark that in this superstring case there is an extra factor of $1/2$, which 
\textit{differs} from the usual result given in \cite{ACNY, CDS, Cheung-Krogh, 
Chu-Ho, Schomerus, Cornalba-Schiappa-1, Seiberg-Witten, Cornalba-2, 
Chu-Zamora}, but is unequivocally determined within the pure spinor formalism.
Given the above equation, the reaction term $\Phi^B_r$ to the 
current $J^B$ must be given by
\begin{eqnarray*}
\Phi_r^B  &=& -\frac{1}{2}  B_{mn}X^m\ c\partial c\ \psi^{n}\ \partial
\phi \partial \xi\ e^{-3\phi}\\
&&
- \frac{1}{2}\ B_{mn}\ c\partial c \partial^{2} c\ \partial 
\xi \partial^{2} \xi\ \psi^{m}\psi^{n}\ e^{-4\phi}\ .
\end{eqnarray*}
\noindent
Note that the relative sign between the two terms is different from the one 
in $\Phi_{(-2)}$. This must be so since $\Phi_{(-2)}$ is BRST closed, whereas 
$Q\Phi_r = -J$. In the $C$--field case one has \textit{an identical current} 
$J$ (up to an overall sign and the replacement $B\to B_{\rm eff}$), and 
therefore the reaction must be identical. Thus, we conclude that
$$
\Phi_c^C + \Phi_r^C = \frac{1}{2}\Phi_{(-2)}(B_{\rm eff})\ ,
$$
\noindent
as we wanted to show.

One would like to develop better these arguments, in a context 
where RR fields appear naturally, in order to fully understand the 
most general cases. This shall be studied next, in the context of the 
pure spinor formalism, which allows for a manifest super Poincar\'e 
covariant quantization of the superstring and where RR fields can be 
studied most simply.

\section{Analysis in the Covariant Formalism}\label{ACF}

Let us start by reviewing the basics of the pure spinor covariant 
formalism, as developed by Berkovits in an extensive list of papers 
\cite{Siegel, Berkovits-1, Berkovits-Vallilo, Berkovits-2, Berkovits-3, 
Berkovits-Chandia-1, Berkovits-4, Berkovits-Chandia-2, Berkovits-Howe, 
Berkovits-5, Berkovits-Chandia-3, Berkovits-Pershin, Berkovits-6}. We 
shall write all the formulae for type IIB string theory, but the 
generalization to type IIA is straightforward.

\subsection{The Underlying Conformal Field Theory}

In this section we recall the conformal field theory underlying the covariant 
formalism in flat space. We shall again use units such that $\alpha' = 2$.
First of all, we have the usual matter part
$$
\frac{1}{2\pi} \int d^2 z\, \partial x^{m}\ol{\partial} x^{n}\, 
g_{mn}\ ,
$$
\noindent
of central charge $+10$.  The spacetime coordinates $x^{m}$ are completed,
to form the IIB ten dimensional superspace, with the addition of the
fermionic chiral spinor coordinates $\theta ^{\alpha}$, 
$\overline{\theta}^{\alpha }$ (see appendix A for spinor conventions).
These fermionic coordinates are promoted to conformal fields of the 
underlying CFT, together with their respective conjugate momenta 
$p_{\alpha}$, $\overline{p}_{\alpha}$, with the action
$$
\frac{1}{\pi} \int d^2z\,\left(p_{\alpha }\overline{\partial }\theta^{\alpha }
+\overline{p}_{\alpha}\partial \overline{\theta }^{\alpha }\right)\, .
$$
\noindent
The fields $\theta ^{\alpha }$ and $p_{\alpha }$ form $16$ fermionic
holomorphic $bc$ systems of conformal dimension $0$ and $1$ respectively,
which contribute $-32$ to the central charge ($-2$ for each of the $16$
pairs). The same holds for the pairs $\overline{\theta }^{\alpha }$, 
$\overline{p}_{\alpha }$ on the anti--holomorphic side. Finally we have the
bosonic pure spinors $\lambda ^{\alpha }$, $\overline{\lambda }^{\alpha }$
together with their conjugate pairs $w_{\alpha }$, $\overline{w}_{\alpha }$
and with an action of the sort
\begin{equation}
``\;\frac{1}{\pi}\int d^2z\, \left( w_{\alpha }\overline{\partial }
\lambda^{\alpha }+
\overline{w}_{\alpha }\partial \overline{\lambda }^{\alpha }\right)\;"\ .
\label{pure_spin_action}
\end{equation}
\noindent
We should think of the pairs $\lambda ^{\alpha }$ and $w_{\alpha }$ as
bosonic $\beta\gamma $ systems of conformal dimension $0$ and $1$. The
reason for the quotation marks around the above action is that the
fields $\lambda $ and $\overline{\lambda }$ are not free, but constrained to 
be pure spinors satisfying the equation
\begin{equation}\label{constraint}
\lambda \gamma ^{m}\lambda = 0 = \overline{\lambda }\gamma ^{m}
\overline{\lambda }\ .
\end{equation}
\noindent
Without this constraint, the action (\ref{pure_spin_action}) would 
contribute $+32$ to the central charge. However, given the above 
constraints, the number of degrees of freedom is reduced and the contribution 
to the central charge is $+22$, as expected for a critical string theory. One 
of the useful results of the covariant formalism is that one can 
actually solve the constraint (\ref{constraint}) using only free fields, in 
which case $\lambda $ and $\overline{\lambda }$ become composite fields and 
the action (\ref{pure_spin_action}) is replaced by a honest action of the 
underlying free fields. We shall not need the details of this construction, 
which can be found in \cite{Berkovits-1}. For future reference, and to fix 
normalizations, let us record some of the basic OPE's of the CFT
\begin{eqnarray*}
\partial x^{m}\left( z\right) \partial x^{n}\left( w\right)  
&\sim &-\frac{g^{mn}}{\left( z-w\right) ^{2}}\ , \\
p_{\alpha }\left( z\right) \,\theta ^{\beta }\left( w\right)  
&\sim &\frac{\delta _{\alpha }{}^{\beta }}{z-w}\ .
\end{eqnarray*}

\subsection{BRST Operator and Closed String Spectrum}

Central to the construction of any string theory is the description of
physical states in terms of BRST cohomology, and the same holds true for the 
covariant formulation \cite{Berkovits-1, Berkovits-2, 
Berkovits-Chandia-2}. More precisely, the conserved BRST current on the 
world--sheet is given, in terms of the basic fields given above, by
$$
J_{BRST} = \left( \lambda d \right) dz - \left( \overline{\lambda}\overline{d}
\right) d\overline{z}\ ,
$$
\noindent
where the spinor current $d_\alpha$ is defined by
$$
\lambda d = \lambda p - \left( \lambda \gamma^m \theta\right) \partial x_m
-\left( \lambda \gamma^m \theta\right) \left( \theta \gamma_m 
\partial\theta\right)
$$
\noindent
and a similar equation holds for $\overline{d}_\alpha$. The BRST operator is 
then given by
$$
Q = Q_L + Q_R = \frac{1}{2\pi i} \oint J_{BRST}\ .
$$
\noindent
The second ingredient in the construction of the physical spectrum is the
grading of the states by ghost number $g = g_L + g_R$. 
This is defined by declaring $g_L(\lambda) = g_R(\ol{\lambda}) = 1$ and 
$g_L(w) = g_R(\ol{w}) = -1$, with the conformal vacuum of ghost number zero 
as usual. With this definition, $g_L(Q_L) = g_R(Q_R) = 1$ as expected. Then, 
the closed string states are identified with the BRST cohomology at total 
ghost number $g=2$. 

The massless spectrum, in the covariant formulation, is particularly simple 
as states are constructed uniquely from the zero modes of the fields and
therefore are represented by functions in type II superspace 
$(x^m, \theta, \ol\theta)$, 
which also depend non--trivially on the pure spinor coordinates 
$\lambda$ and $\ol{\lambda}$. We can then expand any such function in powers 
of $\lambda$ and $\ol{\lambda}$, where the coefficients of the power series 
are functions in superspace, and where terms proportional to 
$\lambda^m \ol{\lambda}^n$ have ghost number $(g_L,g_R) = (m,n)$. The action 
of the BRST operator on these states constructed from the zero modes
is quite simple, since the operators $Q_L$ and $Q_R$ act as differential
operators $\lambda^\alpha D_\alpha$ and $\ol{\lambda}^\alpha\ol{D}_\alpha$ 
given by
\begin{eqnarray*}
Q_L &\to& \lambda D = \lambda\frac{\partial}{\partial \theta} +
\left(\lambda\gamma^m\theta \right)\partial_m\ , \\
Q_R &\to& \ol{\lambda} \ol{D} = \ol{\lambda}\frac{\partial}{\partial 
\ol{\theta}} + \left(\ol{\lambda} \gamma^m\ol{\theta} \right)\partial_m\ .
\end{eqnarray*}

\subsection{Open Strings in the Covariant Formalism}\label{OSCF}

So far we have only described the closed string sector, and we have
consequently considered the CFT as defined on the whole complex plane. We 
shall now introduce open strings by restricting our CFT to the 
upper half plane $\mathbb{H}$ and by imposing boundary conditions
which relate the left--moving to the right--moving modes. We shall work from 
now on with a $Dp$--brane stretched in the directions $0,1,\ldots,p$, and we 
will use indices $a,b,\ldots$, for the directions parallel to the brane and 
indices $i,j,\ldots$, for the transverse ones. For convenience we 
shall place the brane at $x^i = 0$. Also we define
$$
D^m{}_n = \left( \begin{array}{cc}  \delta^a{}_b & 0 \\
0 & -\delta^i{}_j  
\end{array}\right)\, .
$$
\noindent
Then, $D$--brane boundary conditions on $\partial {\mathbb{H}}$ read
\begin{eqnarray}\label{B.C.}
\overline{\partial} x^m &=& D^m{}_n\partial x^m\, , \nonumber \\
\overline{\theta} &=& \gamma^{0}\cdots\gamma^p\theta\, , \nonumber \\
\overline{p} &=& -\gamma^0\cdots\gamma^p p\, , \nonumber \\
\overline{\lambda} &=& \gamma^{0}\cdots\gamma^p\lambda\, .
\end{eqnarray}
\noindent
One can check that the current $J_{BRST}$ does not flow trough the
boundary $\partial {\mathbb{H}}$ and therefore one has a well defined 
conserved BRST operator acting on the open string sector. Moreover, the 
above boundary conditions naturally define the projection operator $\pi$ for 
a closed string state which has a regular OPE with the boundary 
$\partial {\mathbb{H}}$. 

Again we shall focus on the massless sector, so we focus on states built 
from the zero modes of the parallel coordinates $x^a$, the spinor $\theta$ 
and the pure spinor $\lambda$. The massless states are the BRST 
cohomology at ghost number $1$, \textit{i.e.}, with only one $\lambda$.
As in the closed string sector, the BRST operator $Q$ is represented, on 
states built from the zero modes, by a differential operator\footnote{We 
slightly abuse notation by using the same symbol $\lambda D$ for the open and 
closed string sectors, with the hope that it will be clear from context
which is the relevant operator used in the expressions to follow.}
$\lambda^\alpha D_\alpha$ given by
$$
Q \to \lambda D = \lambda\frac{\partial}{\partial \theta} + 
2\left(\lambda\gamma^a\theta \right)\partial_a\, .
$$
\noindent
Note the extra factor of $2$ relative to the analogous expression
for closed strings, and the fact that the differentiation $\partial_a$ is only 
with respect to the parallel coordinates. A massless state of ghost number 
one is then represented by
$$
\Phi  = \lambda^\alpha A_\alpha(x^a,\theta)\, ,
$$
\noindent
where $A_\alpha$ is correctly interpreted as the spinor part of a $U(1)$ 
connection on superspace $(x,\theta)$. The covariant derivatives are 
$\nabla_\alpha = D_\alpha + A_\alpha$ and $\nabla_a = \partial_a + A_a$. One 
may easily check that $\{D_\alpha,D_\beta\} = 4\gamma_{\alpha\beta}^a 
\partial_a$, so it is natural to impose the constraint $\{\nabla_\alpha,
\nabla_\beta\} = 4 \gamma_{\alpha\beta}^a \nabla_a$. Therefore the space 
part $A_a$ of the gauge potential is given by
$$
A_a =\frac{1}{32} \gamma_a^{\alpha\beta} D_\alpha A_\beta\ . 
$$
\noindent
For future reference, we record the expression for the gluon vertex operator
\begin{equation}\label{gluon}
\Phi = 2\left( \lambda \gamma^{a}\theta \right) a_a -\frac{1}{2} ( \lambda 
\gamma_{a}\theta ) ( \theta \gamma^{abc}\theta ) f_{bc} + \cdots\, ,
\end{equation}
\noindent
where $a_a$ is the $U(1)$ gauge field on the brane and where $\cdots$ 
represents terms of higher order in $\theta$. The equation $Q\Phi=0$ implies 
$f=da$ and the Maxwell equations $\partial^a f_{ab} = 0$. Moreover the terms 
in $\cdots$ depend, on--shell, on the derivatives $\partial_a f_{bc}$ of the 
field strength. Therefore, the gluon vertex operator for a constant 
field strength $F_{ab}$ is given explicitly by
\begin{equation}\label{constF}
\Phi = F_{ab} x^a ( \lambda \gamma^{b}\theta )  
-\frac{1}{2} \left( \lambda \gamma_{a}\theta \right)
( \theta \gamma^{abc}\theta ) F_{bc}\ .
\end{equation}

\subsection{Vertex Operators for Constant $B$ and $C$--Fields}

In this section we describe in detail the vertex operators which correspond to 
constant NSNS $B$--field and constant RR $C$--field backgrounds. We 
shall only state the basic facts which we shall need in the remainder of the 
paper, and we refer the reader to the basic references \cite{Berkovits-1, 
Berkovits-3} for further details. Clearly from the point of view of closed 
strings these backgrounds are gauge trivial, and do not affect the closed 
string physics. They will on the other hand affect the dynamics of open 
strings, as we shall prove later.

The gauge trivial vertex operator $\Psi = Q\eta$ for a constant NSNS 
$B$--field $B_{mn}$ is given by
\begin{equation}\label{constB}
\Psi =Q\eta = B_{mn} \left( \lambda \gamma^m \theta\right) 
\left( \ol{\lambda} \gamma^n \ol{\theta}\right)\, ,
\end{equation}
\noindent
where the gauge parameter $\eta$ is
\begin{eqnarray*}
\eta &=& \frac{1}{2}B_{mn}x^m \left( \lambda \gamma^n \theta\right)
-\frac{1}{8} \left( \lambda \gamma_p \theta\right) \left( \theta 
\gamma^{pmn} \theta\right) B_{mn}\\
&&
+\frac{1}{2}B_{mn}x^m \left( \ol\lambda \gamma^n \ol\theta\right) 
-\frac{1}{8} \left( \ol\lambda \gamma_p \ol\theta\right) \left( \ol\theta
\gamma^{pmn} \ol\theta\right) B_{mn}\, .
\end{eqnarray*}
\noindent
The vertex operator for a constant RR potential $C_{\alpha }{}^{\beta }$ 
reads, on the other hand,
\begin{equation}\label{constC}
\Psi = Q\eta = -\frac{1}{4}\left( \lambda \ol{C}\gamma_{m}
\overline{\theta }\right) \left( \overline{\theta }\gamma ^{m}
\overline{\lambda }\right) - \frac{1}{4}
\left(\ol{\lambda}C\gamma_m\theta\right) \left(\theta\gamma^{m}
\lambda\right)\, ,
\end{equation}
\noindent
where
\begin{equation}\label{etaC}
\eta = -\frac{1}{4} \left( \theta \ol{C}\gamma _{m}\overline{\theta }
\right) \left( \overline{\theta }\gamma ^{m}\overline{\lambda }\right)
-\frac{1}{4} \left( \ol{\theta}C \gamma _{m} \theta \right)
\left( \theta \gamma^{m}\lambda \right)\, .
\end{equation}

Let us comment on the form of the vertex operators (\ref{constB}) and 
(\ref{constC}). As in every version of type II string theory, closed string 
vertex operators are tensor products of the left and the right sector, which 
are similar in structure to open string vertex operators. In fact, one 
usually has the correspondence
\begin{eqnarray*}
\Psi_{\mathrm{NSNS}} &\sim& B_{mn} ({\mathrm{Gluon}})^m \otimes 
(\ol{\mathrm{Gluon}})^n\ ,\\
\Psi_{\mathrm{RR}} &\sim& F^{\alpha\beta} ({\mathrm{Gluino}})_\alpha 
\otimes (\ol{\mathrm{Gluino}})_\beta\, ,
\end{eqnarray*}
\noindent
where $2F^{\alpha \beta } =( \overrightarrow{\delsl}C)^{\alpha \beta } + 
( C\overleftarrow{\delsl})^{\alpha\beta}$ is the RR field strength $F=dC$. 
In the covariant formalism one has that
$$
({\mathrm{Gluon}})^m \sim (\lambda \gamma^m \theta)\ ,\ \ \ \ \ \ \ \ \
\ \ \ \ \ \ 
({\mathrm{Gluino}})_\alpha \sim (\lambda \gamma^m \theta)
(\theta \gamma_m)_\alpha\ ,
$$
\noindent
and one might therefore expect that the RR vertex operator is given by
$$
\Psi_{\mathrm{RR}} \sim (\lambda \gamma^m \theta)(\theta \gamma_m F 
\gamma_n\ol{\theta}) (\overline{\theta}\gamma^n \overline{\lambda})\, .
$$
\noindent
The above expression is indeed useful when considering configurations which 
are \textit{not gauge trivial}, since it contains explicitly the field 
strength. On the other hand, following \cite{Berkovits-10, Berkovits-11}
and in analogy with similar results in the RNS formalism \cite{BIANCHI, 
VFPSLR, BVFLPRS}, there is a useful form of the vertex operator
which depends on the RR potential $C$ and which is written as a tensor 
product, 
$$
\Psi_{\mathrm{RR}} \sim C_{\alpha}{}^\beta ({\mathrm{Gauge}})^\alpha 
\otimes (\ol{\mathrm{Gluino}})_\beta + {\mathrm{dual}}\ ,
$$
\noindent
of a right (left) gluino times a left (right) fermion gauge field 
$({\mathrm{Gauge}})^\alpha$, which is nothing but the lowest component of the 
open string gauge potential $\lambda^\alpha A_\alpha$
$$
(\mathrm{Gauge})^\alpha \sim \lambda^\alpha\, .
$$
\noindent
This is indeed expression (\ref{constC}), which is used in the rest of 
the paper.

Let us conclude this subsection with a word on the normalization of the 
states (\ref{gluon}) and (\ref{constB}). The states $\Psi$ and $\Phi$ 
represent, in the language of the covariant formalism, the unintegrated 
vertex operators or the string fields. These states are canonically 
associated to the integrated form of the vertex operators $V_c$ and $V_o$,
which represent the linear deformation to the bulk and boundary sigma 
model respectively,
$$
\int dz\wedge d\overline{z}\  V_c + \oint d\tau\  V_o\, .
$$ 
\noindent
The defining relations for $V_c$ and $V_o$ are \cite{Berkovits-1}
$$
Q_LQ_R V_c = \partial\overline\partial \Psi\, ,\ \ \ \ \ \ \ \ \ \ \
\ \ \ \ \ \ Q V_o = \frac{d}{d\tau} \Phi\, .
$$
\noindent
Given the form of the BRST operator, one may easily check that
$$
V_c = B_{mn} \partial x^m \ol\partial x^n\, , \ \ \ \ \ \ \ \ \ \ \ \ \ 
\ \ \ \ \ V_o = a_m \frac{dx^m}{d\tau}\, ,
$$
\noindent
and therefore that the linear deformation of the sigma model is given by
$$
\int B+\oint a\, .
$$
\noindent
Hence the fields $B_{mn}$ and $a_m$ are canonically normalized. Finally 
note that the same reasoning can be applied to the RNS string, thus showing 
that the states in section \ref{ARNSF} are also correctly normalized.

\subsection{Open Strings in a Constant $B$--Field}\label{OSCBF}

In this section we use the covariant formalism which we have reviewed to 
derive the behavior of open strings in the presence
of a constant NSNS $B$--field. We shall work, for simplicity of exposition, 
with the maximal $D9$--brane, but we will comment on the cases $p<9$ at the 
end of this section. The closed string background is then given by
(\ref{constB}) and, in order to compute the current $J$ and the 
closed string part of the boundary excitation $\Phi_c$ we must bring the 
operators $\Psi$ and $\eta$ to the boundary $\partial {\mathbb {H}}$ of the 
string world--sheet using the boundary conditions described in section 
\ref{OSCF} for $p=9$. We then obtain that
\begin{equation}
J=B_{mn} \left( \lambda \gamma ^{m}\theta \right) \left( \lambda 
\gamma^{n}\theta \right)  = -\frac{1}{96} \left(\theta\gamma_{mnp}\theta\right) 
\left(\lambda \gamma^{mnpqr}\lambda\right)B_{qr}\, ,\label{curr}
\end{equation}
\noindent
and that
$$
\Phi_c =  B_{mn} x^m \left( \lambda \gamma^n \theta\right) -\frac{1}{4} 
\left( \lambda \gamma_p \theta\right) \left( \theta \gamma^{pmn} 
\theta\right) B_{mn}\, .
$$
\noindent
Let us note that the state $\Phi_c$ has the same structure as the open state
(\ref{constF}), but since the relative weights of the two terms are different 
it is therefore \textit{not on--shell} from the open string point of view. In 
fact, in the covariant formalism, one must add a reaction boundary term 
$\Phi_r$ in the case of a constant $B$--field bulk perturbation, differently 
from the bosonic case where the corresponding $\Phi_c$ is already on--shell. 
As we shall show later, in the covariant formalism the $B$ and $C$ bulk 
perturbations look quite similar from the open string perspective, and can be 
treated symmetrically. One must then solve the basic equation
\begin{equation}
Q\Phi _{r}=-J\, ,\ \ \ \ \ \ \ \ \ \ \ \ \ \ \ \ \ \ 
\Phi _{r}=\lambda A(x,\theta)\, .
\label{b1}
\end{equation}
\noindent
In appendix B we describe the general strategy to solve the above source
equation, but in this section we shall solve equation (\ref{b1}) directly. It
will be convenient to consider (\ref{b1}) for a more general current $J$,
which we take to be of the \textit{same form} as in (\ref{curr}), but with a
space dependent field $B$ with vanishing field strength $dB=0$. One can
show that the current $J$ is still BRST closed, \textit{i.e.}, $QJ=0$. This 
is simple to see noting that the combination $\left( \lambda 
\gamma^{m} \theta \right) $ is an odd object which acts in many respects like 
an ordinary differential $dx^{m}$, and that the BRST operator (or, more 
precisely, the operator $\left( \lambda \gamma ^{m}\theta \right) 
\partial _{m}$) acts as the ordinary exterior derivative. To solve (\ref{b1}) 
we expand the gauge field $\lambda A$ as
$$
\lambda A=2 \left( \lambda \gamma ^{m}\theta \right) a_{m}(x) - 
\frac{1}{2} \left( \lambda \gamma _{m}\theta \right) \left( \theta 
\gamma^{mnp}\theta \right) f_{np}(x) + {\mathcal{O}} (\theta^{5}).
$$
\noindent
As we discussed in the previous section, in the absence of a current $J$
the equation $Q(\lambda A)=0$ implies that $f=da$ and that $\partial^{m}
f_{mn}=0$. One the other hand, in the presence of the current (\ref{curr}) 
one obtains
$$
da+\frac{1}{2}B=f\, ,\ \ \ \ \ \ \ \ \ \ \ \ \ \ \partial ^{m}f_{mn}=0\,.
$$

We wish to solve the above equations in the Lorentz gauge 
$\partial^{m}A_{m} = 0$, where we recall that $32\,A_{m}=\gamma_{m}^{\alpha 
\beta }D_{\alpha}A_{\beta}$. At the level of component fields this simply 
implies that $\partial^{m}a_{m}=0$. If we differentiate the equation 
$\partial_{p}a_{n}-\partial _{n}a_{p}+\frac{1}{2}B_{pn}=f_{pn}$ with respect 
to $\partial^{p}$, and we use the gauge condition and the equation of motion 
for $f$ we arrive at the solution for $a_{n}$,
$$
a_{n}=-\frac{1}{2}\frac{1}{\square }\partial^{p}B_{pn}\, ,
$$
\noindent
which may be used to compute $da$:
\begin{eqnarray*}
2 (da)_{mn} &=& \frac{1}{\square} \partial^p \left( -\partial_m B_{pn} + 
\partial_n B_{pm}\right) \\
&=& -\frac{1}{\square} \partial^p \partial_p B_{mn} = -B_{mn}\, .
\end{eqnarray*}
\noindent
This implies that $f$ vanishes, as well as the terms in $\lambda A$ of
higher order in $\theta$. Therefore, the final solution is given by
$$
\Phi _{r}=2 \left( \lambda \gamma ^{m}\theta \right) a_{m}\, ,
$$
\noindent
where
$$
da=-\frac{1}{2} B, \ \ \ \ \ \ \ \ \ \ \ \ \ \ \ \ \ 
\left( \partial \cdot a\right) =0\,.
$$
\noindent
In particular, for a constant field $B$, the solution reads
\begin{equation}\label{react}
\Phi _{r}=-\frac{1}{2}B_{mn}x^{m}\left( \lambda \gamma ^{n}\theta \right) \,.
\end{equation}
\noindent
We have therefore derived that the total open string deformation is given by
$$
\Phi =\Phi_c + \Phi_r = \frac{1}{2}B_{mn} x^m\left( \lambda \gamma^{n}\theta 
\right) - \frac{1}{4} \left( \lambda \gamma_{p}\theta \right)
\left( \theta \gamma^{pmn}\theta \right) B_{mn}\, ,
$$
\noindent
which corresponds to an on--shell gluon with a constant 
field strength $\frac{1}{2}B_{mn}$. Note that this differs from the usually 
assumed result $B_{mn}$. The extra factor of $1/2$, which naturally 
arises in this formalism, is also responsible for the non--vanishing effect 
in the $C$--field case, and will make all the formulae compatible with 
$S$--duality for the $D3$--brane (to be discussed later).
The above argument is very much dependent on the extension of the current
$J$ away from the zero momentum case of constant $B$--field. The reader 
might object that this is arbitrary. On the other hand, we will show in section
6 that equation (\ref{b1}) for constant $B$ is the {\it only equation we 
need to solve whenever considering constant NSNS and RR fields. Therefore,
if we assume that the solution of (\ref{b1}) is given as above, and we
use principle} (A) {\it of the introduction, we are able to treat all the
possible cases at hand. We will then show, again in section 6, that $S$--duality
is non--trivially satisfied using the result for (\ref{b1}) given above. This
is the most compelling proof of the correctness of the results in this section.}

Let us conclude by commenting on the case $p<9$. In this case the form of 
the current $J$ is modified to $J =B_{mn}D^n{}_p \left( \lambda \gamma^{m}
\theta \right) \left( \lambda \gamma^{p}\theta \right)$ since, using the 
projection $\pi$, one shows simply that $\left( \ol{\lambda} \gamma^n 
\ol{\theta}\right) \mapsto  D^n{}_p\left( {\lambda} \gamma^p {\theta}\right)$.
As usual one must distinguish between a $B$--field transverse and 
parallel to the brane. The parallel case is identical to the case above, 
whereas the transverse case can be analyzed with a similar computation. In 
this latter case though one finds that the resulting deformation of the open 
string $\Phi$ is BRST trivial, and therefore the $B$--field induces no 
physical effect on the open strings. This is in agreement with the usual 
result described in \cite{Seiberg-Witten}.

\subsection{$D$--Branes in an Arbitrary Constant $C$--Field}

We are ready to consider the central problem of this paper, 
\textit{i.e.}, the behavior of open strings in the presence of a constant RR 
potential. In the first part of this section we shall discuss in detail 
the $D9$--brane case, leaving the analysis of the lower dimensional branes
to the second part of this section.

\subsubsection{The $D9$--Brane}

Using the $D9$--brane boundary conditions we can deduce that the current 
$J$ is
\begin{eqnarray*}
J &=& - \frac{1}{4}\left( \lambda [C+\ol{C}]\gamma _m\theta\right) 
\left(\theta \gamma ^m\lambda\right)\\
&=& \frac{1}{96}\left(\theta \gamma_{mnp}\theta\right) \left(\lambda 
[C+\ol{C}]\gamma^{mnp}\lambda\right) ,
\end{eqnarray*}
\noindent
and that $\Phi_c$ is
$$
\Phi_c = -\frac{1}{4} \left(\lambda\gamma^m\theta\right) 
\left(\theta[C+\ol{C}]\gamma^m\theta\right)\, .
$$
\noindent
The combination $C+\ol{C}$ is non--zero only for a $q$--form potential 
$C^{(q)}$ with $q=0,4,8$. The case $q=0$ gives immediately $J=\Phi_c=0$, 
which implies that the open string physics is unaltered. Let us consider 
the other two cases starting with the case $q=8=p-1$, which should lead to a
Moyal noncommutativity on the brane. Using the fact that
$$
\frac{1}{8!} C^{(8)}_{m_1\cdots m_8} \left(\gamma^{m_1\cdots m_8}
\right)_\alpha{}^\beta = -\frac{1}{2!} (\star C^{(8)})_{mn} \left(\gamma^{mn}
\right)_\alpha{}^\beta\, ,
$$
\noindent
we deduce that
\begin{eqnarray*}
J &=&  -\frac{1}{96}\left(\theta \gamma_{mnp}\theta\right) 
\left(\lambda\gamma^{mnpqr}\lambda\right) (\star C^{(8)})_{qr}\, ,\\
\Phi_c &=& \frac{1}{4} \left(\lambda\gamma^m\theta\right) \left(\theta
\gamma^{mnp}\theta\right)(\star C^{(8)})_{np}\, .
\end{eqnarray*}
\noindent
We note that $J$ has the same form as (\ref{curr}) with the replacement of 
$B$ with $\star C^{(8)}$. This means that we can use the results of the 
previous section to deduce that $\Phi_r$ is given by the expression 
(\ref{react}), again up to the replacement $B \mapsto \star C^{(8)}$. This 
then implies that the total open string deformation is given by
$$
\Phi =\Phi_c + \Phi_r = \frac{1}{2}(B_{\rm{eff}})_{mn} x^m\left( \lambda 
\gamma^{n}\theta \right) - \frac{1}{4} \left( \lambda \gamma_{p}\theta 
\right) \left( \theta \gamma^{pmn}\theta \right) 
(B_{\rm{eff}})_{mn}\, ,
$$
\noindent
where
$$
B_{\rm{eff}} = -\star C^{(8)}\, .
$$
This shows that the effect of a $C^{(8)}$--field on a $D9$--brane is 
equivalent to an effective $B$--field, as claimed at the beginning of the 
paper.

Next let us consider the case $q=4$. This leads to
\begin{eqnarray*}
J &=&\frac{1}{1152}\,C_{q_{1}\cdots q_{4}}\,\left( \theta \gamma
_{mnp}\theta \right) \left( \lambda \gamma ^{q_{1}\cdots q_{4}}\gamma
^{mnp}\lambda \right) \,, \\
\Phi _{c} &=&\frac{1}{12}C_{q_{1}\cdots q_{4}}\,\left( \lambda \gamma
^{q_{1}}\theta \right) \left( \theta \gamma ^{q_{2}\cdots q_{4}}\theta
\right) \,.
\end{eqnarray*}
One may follow the general procedure, and solve the reaction equation $Q\Phi
_{r}=-J$ using the general techniques of appendix B. On the other hand, in
this case, there is a short--cut which allows us to avoid the use of the
general machinery. As for the case $q=8$, the general reaction term can only
be a linear combination of a term linear in $\theta $ and a term cubic in $%
\theta $, so that the final boundary deformation is
\begin{equation*}
\Phi =\alpha \,C_{q_{1}\cdots q_{4}}\left[ x^{q_{1}}\,\left( \lambda \gamma
^{q_{2}\cdots q_{4}}\theta \right) +\left( \lambda \gamma ^{q_{1}}\theta
\right) \left( \theta \gamma ^{q_{2}\cdots q_{4}}\theta \right) \right] \,,
\end{equation*}
where $\alpha $ is a constant, and where the relative coefficient between
the two terms is fixed by the on--shell condition $Q\Phi =0$. The precise
value of the constant $\alpha $ can be determined by explicitly solving the
equation $Q\Phi _{r}=-J$. On the other hand we will not need the value of $%
\alpha $ since the deformation $\Phi $ is \textit{BRST trivial }for any
value of $\alpha $, since $\Phi =Q\Lambda $, where
\begin{equation*}
\Lambda =\frac{1}{2}\alpha \,C_{q_{1}\cdots q_{4}}x^{q_{1}}\,\left( \theta
\gamma ^{q_{2}\cdots q_{4}}\theta \right) \,.
\end{equation*}
Therefore, a $C^{\left( 4\right) }$ RR potential has no effect on the 
$D9$--brane open string dynamics.

\subsubsection{$Dp$--Branes for $p<9$ with Parallel RR Potential}

We now move to the general case of arbitrary $p$. Using the appropriate 
boundary conditions we deduce that
\begin{eqnarray*}
J &=& -\frac{1}{4} \left(\lambda [\gamma^{p \cdots 0} C\gamma_m+
D^n{}_{m}\ol{C} \gamma_n \gamma^{0\cdots p} 
]\theta\right)\left(\theta\gamma^m\lambda\right)\\
&=&\frac{1}{96} \left(\theta\gamma_{mnp}\theta\right)
\left(\lambda [\gamma^{p \cdots 0} C-\ol{C}  \gamma^{0\cdots p}]
\gamma^{mnp}\lambda\right) ,
\end{eqnarray*} 
\noindent
and similarly
$$
\Phi_c = -\frac{1}{4} \left(\lambda\gamma^m\theta\right) \left(\theta
[\gamma^{p \cdots 0} C-\ol{C}  \gamma^{0\cdots p}]
\gamma_m\theta\right)\, .
$$
\noindent
As before, we will assume that we have a $C^{(q)}$--form potential, and we 
will start to analyze the case of a form potential \textit{parallel} to the 
brane. It is tedious but easy to show that
\begin{eqnarray*}
\gamma^{p\cdots 0} C^{(q)} &=&  \left( -\right)^{\frac{(p-q+1)(p-q)}{2}} 
\star C^{(q)}\, ,\\
-\ol{C}^{(q)} \gamma^{0\cdots p} &=&  - \star C^{(q)}\, ,
\end{eqnarray*}
\noindent
where $\star$ is the Hodge dual \textit{on the brane world--volume}. 
Therefore the sum $\gamma^{p \cdots 0} C-\ol{C}  \gamma^{0\cdots p}$ vanishes 
if $p-q = 3+4n$ and is equal to $-2(\star C^{(q)})$ if $p-q = 1+4n$. The two 
relevant cases are therefore $q=p-1$ and $q=p-5$. If 
$q=p-1$ one can show, using a reasoning analogous to the one used for the 
$D9$ case, that the effective deformation has a physical effect and 
corresponds to a gluon with a constant field strength
$$
B_{\rm{eff}} = -\star C^{p-1}\, .
$$
In the case $q=p-5$ we obtain, as for the case $p=9$ of the last subsection,
a gauge trivial deformation on the boundary which does not effect the open
string dynamics.

\subsubsection{$Dp$--Branes for $p<9$ with General RR Potential}

In this last subsection we are going to study some examples of the most 
general case of a $Dp$--brane immersed in a constant RR field which is not 
necessarily parallel to the brane. In particular we shall be interested in 
the case where the form
$$
\gamma^{p\cdots 0} C
$$
\noindent
is a two--form\footnote{One has a vanishing contribution when 
$\gamma^{p\cdots 0} C$ is a $0$, $4$ or an $8$--form. The
other non--trivial contributions, which we are not considering, come from a
$6$ or a $10$--form.}. More precisely we will consider the potential
$$
C^{(q+k)} = C_\parallel^{(q)} \wedge C_\perp^{(k)}\, ,
$$
\noindent
where $C_\parallel^{(q)}$ and $C_\perp^{(k)}$ are, respectively, the parallel and 
perpendicular $q$--form and $k$--form parts of the RR 
$(q+k)$--form potential $C^{(q+k)}$. Since the matrix $\gamma^{p\cdots 0}$
effectively computes the Hodge dual $\star$ on the brane world--volume, we 
are requiring that $\star \left( C_\parallel^{(q)} \wedge  
C_\perp^{(k)} \right)$ is a two--form. We have then three possibilities for 
the values of the pair $(q,k)$. First we have $(p-1,0)$ which is the case 
studied in the previous section. The other two possibilities are $(p,1)$ 
and $(p+1,2)$. In all three cases we define the two--form
$$
\Omega = (-)^\frac{(p-q+1)(p-q)}{2} \star \left( C_\parallel^{(q)} \wedge  
C_\perp^{(k)} \right) = \frac{1}{2}\Omega_{mn} dx^m\wedge dx^n\, ,
$$
\noindent
which can have components both parallel and transverse to the brane 
world--volume. Then we have that
$$
J = \frac{1}{96} \left(\theta\gamma_{mnp}\theta\right)
\left(\lambda
\gamma^{mnpqr}\lambda\right)\Omega_{qr}\, ,
$$ 
\noindent
and similarly
$$
\Phi_c = -\frac{1}{4} \left(\lambda\gamma_m\theta\right) \left(\theta
\gamma^{mnp}\theta\right)\Omega_{np}\, .
$$
\noindent
We consider again the source equation $Q\Phi_r =-J$ with $\Phi_r$ given in 
general by
\begin{eqnarray*}
\Phi_r &=&2\left( \lambda \gamma ^{a}\theta \right) a_{a} + 2\left( 
\lambda \gamma ^{i}\theta \right) z_{i}\\
&&-\frac{1}{2} ( \lambda \gamma _{m}\theta ) ( \theta \gamma^{mab}
\theta ) f_{ab} - ( \lambda \gamma _{m}\theta )( \theta \gamma^{mai}
\theta ) g_{ai}\\
&&-\frac{1}{2} ( \lambda \gamma _{m}\theta )( \theta \gamma^{mij}
\theta ) h_{ij} + \cdots\, .
\end{eqnarray*}
\noindent
The fields $a_a$ and $z_i$ are the gluon field and the transverse scalars 
of the brane world--volume, respectively. The source equation then reads
\begin{eqnarray*}
(da)_{ab} - \frac{1}{2}\Omega_{ab} = f_{ab}\, ,&&\\
\partial_a z_i -\frac{1}{2}\Omega_{ai} = g_{ai}\, ,&& \\
-\frac{1}{2}\Omega_{ij} = h_{ij}\, ,&&
\end{eqnarray*}
\noindent
together with
$$
\partial^a f_{ab} = \partial^a g_{ai} = 0\, .
$$
\noindent
Following a reasoning similar to the one used in section \ref{OSCF}, or 
using the general solution technique of appendix B, we conclude that
$$
a_b = \frac{1}{4} x^a \Omega_{ab}\, , \ \ \ \ \ \ \ \ \ \ \ \ \ \ \ 
z_i = \frac{1}{2}x^a \Omega_{ai}\, .
$$
\noindent
Therefore the total open string deformation $\Phi = \Phi_c + \Phi_r$ is 
given by
\begin{eqnarray}\label{general}
\Phi &=& \frac{1}{2} \Omega_{ab} x^a ( \lambda \gamma ^{b}\theta) 
-\frac{1}{4} (\lambda\gamma_m\theta)(\theta
\gamma^{mab}\theta)\Omega_{ab}\\
&&+\Omega_{ai} x^a ( \lambda \gamma ^{i}\theta ) 
-\frac{1}{2} (\lambda\gamma_m\theta) (\theta
\gamma^{mai}\theta)\Omega_{ai}\, .\notag
\end{eqnarray}
\noindent
Let us comment on this result. The first line, which comes from the case 
$(q,k) = (p-1,0)$, represents a gluon field with constant field strength, 
and is the result of the last section. The case $(q,k)= (p+1,2)$ gives no 
deformation since, in this case, $\Omega$ only has components $\Omega_{ij}$
transverse to the brane. The second line of (\ref{general}), coming from 
the case $(q,k) = (p,1)$ is quite interesting and different from the 
$B$--field case. It represents a transverse displacement of the brane 
$z^i = \frac{1}{2}x^a \Omega_{ai}$. Recall that in the $B$--field case, a 
field $B_{ai}$ gives no deformation. This can be seen easily by noting that 
the current 
$$
J=B_{mn} D^n{}_p\left( \lambda \gamma ^{m}\theta \right) \left( \lambda 
\gamma^{p}\theta \right)
$$
\noindent
vanishes since $B_{an}D^{n}{}_{i} = B_{in}D^{n}{}_{a}$. The same holds for 
the corresponding state $\Phi_c$.

\section{The Non--Linear Open String Parameters and $S$--Duality}\label{NLOSP}

In the previous section we have analyzed the effect on the open strings to
leading order in the bulk perturbation. In the following we wish to address
the complete solution to the problem, by extending the previous results 
to general boundary conditions on the brane and, as we describe below, to
all orders in the closed string vertex operators.

Let us describe the general strategy. We shall concentrate first on the 
$D9$--brane case, on which we introduce a constant $U\left( 1\right) $ gauge
field strength $F_{mn}$. As is well known \cite{ACNY}, a
constant field strength is described by an exact boundary CFT, in which the
boundary conditions (\ref{B.C.}) are altered by introducing a finite Lorentz 
rotation between the left and right--movers 
\begin{eqnarray}
\left( g-F\right) _{mn}\overline{\partial }x^{n} &=&\left( g+F\right) _{mn}
{\partial }x^{n}\,, \notag \\
\overline{\theta } &=&e^{\omega \cdot \gamma }{\theta \,}, \label{Fboundary} \\
\overline{\lambda } &=&e^{\omega \cdot \gamma }{\lambda }\,. \notag
\end{eqnarray}
\noindent
The $SO\left( 1,9\right) $ rotation matrix is parametrized by the field
strength $F_{mn}$ on the brane and is given explicitly by 
\begin{equation*}
{\left( e^{2\omega }\right)^{m}}_{n} = {\left( \frac{1+\frac{1}{g}F}
{1-\frac{1}{g}F} \right)^{m}}_{n}\ ,
\end{equation*}
\noindent
where we use the short--hand notation $\omega \cdot \gamma =
\frac{1}{2} \omega _{mn}\gamma ^{mn}$. A bulk insertion of a gauge trivial 
closed string vertex operator will have the effect of changing the rotation
matrix, and therefore the two--form $F$. Therefore, we will be able to write 
down, as discussed in the introduction, 
an explicit equation for the variation $\delta F$ of $F$, of the 
schematic form 
\begin{equation}
\delta F=\delta F\left( F,B,C\right)\,,   \label{deform}
\end{equation}
\noindent
which will yield the corrections to $F$ due to the constant fields $B$ and 
$C$. The function $\delta F\left( F,B,C\right) $ will be linear in the bulk
fields $B,C$ (treated as small perturbations) but will contain arbitrary
powers of $F$. The results from the previous sections imply that 
\begin{equation*}
\delta F=\frac{1}{2}\left( B-\star C^{\left( 8\right) }\right) +\mathcal{O}
\left( F\right) .
\end{equation*}
\noindent
In this section we shall describe how to compute the 
$\mathcal{O}\left( F\right)$ terms, and we will explicitly write down the 
first terms linear in $F$. Moreover we shall extend these results to the 
$D3$--brane case, and we will show how the explicit form of $\delta F$ is 
compatible with $S$--duality. This is a strong check that our method is
compatible with the requirements (A) and (B) in the introduction. Moreover, notice that
this check of $S$--duality does not require the analysis of string theory 
at strong coupling, but only requires knowledge of the underlying conformal field theory. 
Finally one can, in principle, integrate equation (\ref{deform}) by considering a 
large closed string background as a sum of infinitesimal deformations. One 
then obtains the full deformation $F$ as a non--linear function of the 
background fields $B,C$. One may compute accordingly the open string 
parameters $G$ and $\Theta $ by the usual equation \cite{ACNY, 
Schomerus, Seiberg-Witten}
\begin{equation*}
\frac{1}{G}+\frac{\Theta }{2\pi \alpha ^{\prime }}=\frac{1}{g+F}\text{ }.
\end{equation*}

In order to compute $\delta F$ we first note that, given the general
boundary conditions (\ref{Fboundary}), the boundary BRST operator differs
from the $F=0$ case since the part coming from $Q_{R}$ is rotated with
respect to the part coming from $Q_{L}$. More specifically, for massless
states built only from the zero modes of the world--sheet fields, the BRST
operator reads 
\begin{eqnarray*}
Q &\rightarrow &\lambda \frac{\partial }{\partial \theta }+\left[ \delta
^{m}{}_{n}+\left( e^{2\omega }\right) ^{m}{}_{n}\right] \left( \lambda
\gamma ^{n}\theta \right) \frac{\partial }{\partial x^{m}} \\
&=&\lambda \frac{\partial }{\partial \theta }+2\left( \lambda \gamma
^{m}\theta \right) \frac{\partial }{\partial y^{m}}\,,
\end{eqnarray*}
\noindent
where the new coordinates $y^{m}$ are defined via
\begin{equation*}
y^{m} = {\left( 1-\frac{1}{g}F \right)^{m}}_{n} x^{n}\ .
\end{equation*}
\noindent
It is then clear from the general arguments of the previous sections that
the on--shell constant field strength open string field is of the form 
\begin{equation}
\Phi =\left[ \frac{1}{2}y^{m}\left( \lambda \gamma ^{n}\theta \right) 
-\frac{1}{4}\left( \lambda \gamma _{r}\theta \right) \left( \theta 
\gamma^{mnr}\theta \right) \right] \Omega _{mn}\,,  \label{bfieldcase}
\end{equation}
\noindent
where $\Omega$ is a constant two--form. The above string field is 
associated, in a canonical way, to an integrated vertex operator $V$ which
must be 
\begin{equation*}
V=\frac{1}{2}\delta F_{mn}x^{m}dx^{n}\, ,
\end{equation*}
\noindent
in order to produce, when exponentiated in the sigma model, the correct
change in the boundary conditions. The relation between $V$ and $\Phi $ is
given, like in bosonic string theory, by the equation \cite{Berkovits-1}
\begin{equation*}
QV=d\Phi\, ,
\end{equation*}
\noindent
which, in turn, determines the relation 
\begin{equation}
\delta F=\frac{1}{2}\left( 1+F\frac{1}{g}\right) \Omega \left( 
1-\frac{1}{g}F\right)  \label{deltaF}
\end{equation}
\noindent
between $\Omega $ and $\delta F$. Therefore, as long as the boundary
deformation is of the form (\ref{bfieldcase}), the deformation 
$\delta F$ is given by the equation above. This is the main equation 
to integrate in order to determine the full deformation as a 
non--linear function of the closed string background fields. In the 
following, we shall study the integration of this differential 
equation to linear order in $F$.

Let us discuss the $B$--field case first. The boundary current 
$J$ is given by 
\begin{equation*}
J=B_{mp}\left( e^{2\omega }\right) ^{p}{}_{n}\left( \lambda \gamma
^{m}\theta \right) \left( \lambda \gamma ^{n}\theta \right) ,
\end{equation*}
\noindent
and therefore the reaction term $\Phi _{r}$ is 
\begin{equation*}
\Phi _{r}=-\frac{1}{4}\left[ B_{mp}\left( e^{2\omega }\right)
^{p}{}_{n}-B_{np}\left( e^{2\omega }\right) ^{p}{}_{m}\right] \,y^{m}\left(
\lambda \gamma ^{n}\theta \right) .
\end{equation*}
\noindent
Using crucially requirement (A) of the introduction,
it is straightforward to compute the closed part $\Phi _{c}=\pi \left(
\eta \right) $ of the full open string deformation $\Phi =\Phi _{c}+\Phi
_{r}$, which is given by 
\begin{eqnarray*}
\Phi _{c} &=&-\frac{1}{2}\left. \left( \frac{1}{1-\frac{1}{g}F}\right)
^{p}\right. _{m}B_{pq}\,\left[ \delta ^{q}{}_{n}+\left( e^{2\omega }\right)
^{q}{}_{n}\right] y^{m}\left( \lambda \gamma ^{n}\theta \right)  \\
&&-\frac{1}{8}\left( \lambda \gamma _{r}\theta \right) \left( \theta \gamma
^{mnr}\theta \right) B_{pq}\left[ \delta ^{p}{}_{m}\delta ^{q}{}_{n}+\left(
e^{2\omega }\right) ^{p}{}_{m}\left( e^{2\omega }\right) ^{q}{}_{n}\right] .
\end{eqnarray*}
\noindent
Adding the two contributions we obtain $\Phi$ of the form 
(\ref{bfieldcase}), where 
\begin{equation*}
\Omega =\frac{1}{1+F\frac{1}{g}}\left( B-F\frac{1}{g}B\frac{1}{g}F\right) 
\frac{1}{1-\frac{1}{g}F}\,.
\end{equation*}
\noindent
Therefore we conclude that, for the $B$--field case, one has\footnote{
 Let us note that the naive integration of equation (\ref{bfieldtotal}) yields
\begin{equation*}
F=g\tanh \left( \frac{1}{2}\frac{1}{g}B\right) .
\end{equation*}
It is interesting to observe that, for a time--like $B$--field $%
B=b\,dx^{0}\wedge dx^{1}$, the induced $F$--field becomes $F=\tanh \left( 
\frac{1}{2}b\right) dx^{0}\wedge dx^{1}$, and therefore never generates a
super--critical electric field on the brane.
}
\begin{equation}
\delta F=\frac{1}{2}\left( B-F\frac{1}{g}B\frac{1}{g}F\right) .
\label{bfieldtotal}
\end{equation}

We now turn to the analysis of $\delta F$ in the presence of the RR fields 
$C$. Starting with the explicit form of $\eta $ in the $C$--field case 
(\ref{etaC}), one can use the boundary conditions above to conclude that 
\begin{equation*}
\Phi _{c}=-\frac{1}{4}\left( \lambda \gamma _{r}\theta \right) \left( 
\theta \left[ e^{-\omega \cdot \gamma }C+\overline{C}e^{\omega \cdot 
\gamma }\right] \gamma ^{r}\theta \right) ,
\end{equation*}
\noindent
where we have used the fact that $e^{-\omega \cdot \gamma}
\gamma^{m} e^{\omega \cdot \gamma} = \left( e^{2\omega }\right)^{m}{}_{n}
\gamma^{n}$. One can then conclude, without computations, that the form of 
$\Phi = \Phi_{c} + \Phi_{r}$ in the $C$--field case is again identical to 
(\ref{bfieldcase}), where $\Omega $ is now defined by 
\begin{equation}
\frac{1}{2}\Omega _{mn}\gamma ^{mn}=\frac{1}{2}\left[ e^{-\omega \cdot
\gamma }C+\overline{C}e^{\omega \cdot \gamma }\right] _{2\text{--form}}\,.
\label{omega2}
\end{equation}
\noindent
This follows simply from the fact that $\Phi _{r}$ is linear in $\theta $
and does not affect the $\theta ^{3\,}$terms in $\Phi $, which, again using
requirement (A), come uniquely
from $\Phi _{c}$. Equation (\ref{omega2}), together with (\ref{deltaF}),
determines $\delta F$ as a function of the background $C$--field, to all
orders in $F$. Let us compute explicitly the corrections to $\delta F$ which
are linear in $F$. We start by expanding $\Omega $ (recalling that $\omega
=F+\frac{1}{3}F^{3}+\cdots $) to linear order in $F$, and obtain 
\begin{equation*}
\frac{1}{2}\Omega _{mn}\gamma ^{mn}=-\frac{1}{2}\star C_{mn}^{\left(
8\right) }\gamma ^{mn}-\frac{1}{4}F_{pq}\left[ \gamma ^{pq}C-\overline{C}%
\gamma ^{pq}\right] _{2\text{--form}}\,.
\end{equation*}
\noindent
Concentrating on the terms linear in $F$ we have 
\begin{equation*}
-\frac{1}{4}F_{pq}\sum_{k=0,4,8}\left[ \gamma ^{pq},C^{\left( k\right) }
\right] _{2\text{--form}}-\frac{1}{4}F_{pq}\,\sum_{k=2,6,10}\left\{ \gamma
^{pq},C^{\left( k\right) }\right\} _{2\text{--form}}\, ,
\end{equation*}
\noindent
which yields for $\Omega$ 
\begin{eqnarray*}
&&-F\wedge \star C^{\left( 10\right) }-8\star \left( \frac{1}{8!}
C_{m_{1}\cdots m_{7}n}^{\left( 8\right) }F^{n}{}_{m_{8}}dx^{m_{1}}\cdots
dx^{m_{8}}\right) + \\
&&+\star \left( F\wedge C^{\left( 6\right) }\right) \,.
\end{eqnarray*}
\noindent
Contributions for $k<6$ vanish. Moving to the equation for $\delta F$, 
one can expand (\ref{deltaF}) to linear order in $F$ in order to 
finally obtain 
\begin{eqnarray*}
\delta F &=&-\frac{1}{2}\star C^{(8)}-\frac{1}{2}F\wedge \star C^{\left(
10\right) }+\frac{1}{2}\star \left( F\wedge C^{\left( 6\right) }\right) + \\
&&-\frac{1}{2}\frac{1}{7!}\star \left( C_{m_{1}\cdots m_{7}n}^{\left(
8\right) }F^{n}{}_{m_{8}}dx^{m_{1}}\cdots dx^{m_{8}}\right)  \\
&&+\frac{1}{2}\left( \star C_{mp}^{\left( 8\right) }\,F^{p}{}_{n}\right)
dx^{m}dx^{n}+\mathcal{O}\left( F^{2}\right) \,.
\end{eqnarray*}
\noindent
The last two lines of the equation above actually sum to zero. This is easily shown by
considering the dual identity
$$
\frac{1}{2}\frac{1}{7!} \left( C_{m_{1}\cdots m_{7}n}^{\left(
8\right) }F^{n}{}_{m_{8}}dx^{m_{1}}\cdots dx^{m_{8}}\right) 
+\frac{1}{2}\star\left[\left( \star C_{mp}^{\left( 8\right) }\,F^{p}{}_{n}\right)
dx^{m}dx^{n}\right] = 0\, .
$$
\noindent
If one writes the second term of the equation above in components, using the 
$\varepsilon$--symbol for the duals, and one uses the usual expression for 
the contraction of two $\varepsilon$--symbols in terms of $\delta$'s, one 
recovers the first term with the opposite sign. We conclude that the total 
deformation $\delta F$, to linear order in $F$, is given by
$$
\delta F = -\frac{1}{2}\star C^{(8)}-\frac{1}{2}F\wedge \star C^{\left(
10\right) }+\frac{1}{2}\star \left( F\wedge C^{\left( 6\right) }\right) 
+ \mathcal{O}\left( F^{2}\right) \,.
$$

\subsection{The $D3$--Brane Case}

Let us review the results of the last subsection as they apply to 
the $D3$--brane case. We consider, for simplicity, only fields $B$, $C$ 
\textit{along the brane world--volume}. The results in this case can 
be naively obtained using $T$--duality, as we show below. 
The correct boundary conditions for 
the world--sheet fields are clearly 
\begin{eqnarray*}
\left( g-F\right) _{ab}\overline{\partial }x^{b} &=&\left( g+F\right)_{ab}
{\partial }x^{b}\,, \\
\overline{\theta } &=&e^{\omega \cdot \gamma }\,\gamma ^{9}\cdots \gamma
^{4}\,{\theta \,}, \\
\overline{\lambda } &=&e^{\omega \cdot \gamma }\,\gamma ^{9}\cdots \gamma
^{4}\,{\lambda }\,,
\end{eqnarray*}
\noindent
where $F_{ab}$ is also along the brane directions $0,\ldots,3$. The result 
\begin{equation*}
\delta F=\frac{1}{2}\left( B-F\frac{1}{g}B\frac{1}{g}F\right) 
\end{equation*}
\noindent
in the case of a small $B$--field bulk deformation goes through without
change. The $C$--field case is only slightly more complex. Following the
usual procedure, and using the boundary conditions above we conclude that 
\begin{equation*}
\Omega _{mn}\gamma ^{mn}=\left. e^{-\omega \cdot \gamma }\,C\,\gamma
^{4}\cdots \gamma ^{9}-\overline{C}\,\gamma ^{4}\cdots \gamma
^{9}\,e^{\omega \cdot \gamma }\right| _{2\text{--form}}\,.
\end{equation*}
\noindent
Given $\Omega $, one may then compute $\delta F$ using (\ref{deltaF}). It is
clear from the equation above that, in the presence of fields $C^{\left(
k\right) }$ parallel to the brane, with $k=0,2,4$, one may immediately use
the results obtained for the $D9$--brane case, with the $6,8,10$ RR--forms
now given by $C^{\left( k\right) }\wedge dx^{4}\wedge \cdots \wedge dx^{9}$.
We may therefore readily write down the result for $\delta F$ to linear
order in $F$ as 
\begin{equation}
\delta F=\frac{1}{2}\left( B-\star C^{(2)}\right) -\frac{1}{2}F\wedge \star
C^{\left( 4\right) }+\frac{1}{2}\star F\wedge C^{\left( 0\right) } + 
{\mathcal{O}}(F^{2}) \, ,
\label{finalD3}
\end{equation}
\noindent
where $\star $ is, as always, the Hodge dual \textit{along the brane
world--volume}.

\subsection{A Perturbative Check of $S$--Duality}

To conclude this section, we wish to check the compatibility of the 
above result for the $D3$--brane with $S$--duality. We set $C^{\left(0\right)}=0$, 
since under $S$--duality $C^{\left( 0\right) }$ mixes with the coupling 
constant and we wish on the other hand to make statements purely in CFT, 
at zero coupling. Let us first recall that, under $S$--duality, the bulk 
fields $B$, $C^{\left( 2 \right)}$ and $C^{\left( 4 \right)}$ are mapped to 
$C^{\left( 2 \right)}$, $-B$ and $C^{\left( 4 \right)}$, respectively. 
Moreover, for vanishing closed string fields the field $F$ is mapped 
(to leading order in derivatives, but to 
all orders in $F$) to the two--form \cite{GibbRash,Tsey,GreenGutp}
\begin{equation*}
S\left( F\right) =\sqrt{-\det \left( g+F\right) }\star \left( \frac{1}
{1+F\frac{1}{g}}F\frac{1}{1-\frac{1}{g}F}\right) =\star F+\mathcal{O}
\left(F^{3}\right) \,.
\end{equation*}
\noindent
Consider then a $D3$--brane with a background $U(1)$ gauge field strength
$F$, and let us turn on some closed string fields, collectively denoted by $\Psi$.
We might also consider the $S$--dual process, where we start with the $D3$--brane
with field strength $S\left( F\right)$ and we turn on closed fields 
$S\left( \Psi \right) $. In the first case we arrive, after backreaction
of the open strings, to a variation of the field strength from $F$ to
$$
F' = F+\delta F\left( F,\Psi \right)\, .
$$
In the second case we arrive at the field strength
$$
\widetilde{F}' = S\left( F\right)
+\delta F\left( S\left( F\right),\, S\left( \Psi \right) \right)\, .
$$
Then, $S$--duality implies that $S\left( F' \right) = \widetilde{F}'$ or,
infinitesimally, that
\begin{equation}
\delta F\left( S\left( F\right) ,S\left( \Psi \right) \right) =\delta
F\left( F,\Psi \right) \frac{\partial S}{\partial F}\left( F\right) .
\label{fullSduality}
\end{equation}
\noindent
To linear order in $F$, the above equation reads\thinspace $\delta F\left(
\star F,S\left( \Psi \right) \right) =\star \delta F\left( F,\Psi \right)$,
so that the following relation must hold
\begin{equation*}
\star \delta F\left( F,B,C^{\left( 2\right) },C^{\left( 4\right) }\right)
=\delta F\left( \star F,C^{\left( 2\right) },-B,C^{\left( 4\right) }\right) .
\end{equation*}
\noindent
Clearly equation (\ref{finalD3}) satisfies the above requirement, 
therefore being compatible with $S$--duality.

Next we wish to extend the check of equation (\ref{fullSduality}) to all
orders in $F$. For simplicity of exposition, we shall work from now on with 
$C^{\left( 4\right) }$ set to zero, and we will concentrate on the following
illustrative example 
\begin{equation}
F=f\,dx^{2}\wedge dx^{3}\,,\ \ \ \ \ \ \ \ \ \ \ \ \ \ \ \ \ \ \ \ \ \ \ \ \
\ \ C^{\left( 2\right) }=\epsilon \,dx^{1}\wedge dx^{2}\,,
\label{s1}
\end{equation}
\noindent
where $f$ is an arbitrary constant and where $\epsilon$ is an infinitesimal
parameter. For later convenience we also introduce the angle $\theta$
defined by $\tan \theta =f$ or, equivalently, by 
\begin{equation*}
e^{2i\theta }=\frac{1+if}{1-if}\,.
\end{equation*}
\noindent
Under $S$--duality the field strength $F$ is mapped to 
\begin{equation}
S\left( F\right) =\sin \theta \,dx^{0}\wedge dx^{1}\,.
\label{s2}
\end{equation}
\noindent
Let us introduce the constant two--form $\kappa_{ab}$ with non--zero
entries $\kappa _{23}=-\kappa _{32}=1$. Then $F_{ab}=f\,\,\kappa _{ab}$ and
one notes that, since $\kappa ^{2}=-1$,  
\begin{equation*}
\omega =F+\frac{1}{3}F^{3}+\cdots =\kappa \left( f-\frac{1}{3}f^{3}+\cdots
\right) =\kappa \frac{1}{2i}\ln \left( \frac{1+if}{1-if}\right) =\kappa
\theta\, ,
\end{equation*}
\noindent
and that
\begin{eqnarray*}
e^{2\omega } &=&\cos \left( 2\theta \right) +\kappa \sin \left( 2\theta
\right) \,, \\
e^{\pm \omega \cdot \gamma } &=&e^{\pm \theta \gamma ^{2}\gamma ^{3}}=
\cos\theta\,\pm\,\sin\theta\,\gamma ^{2}\gamma^{3}\,.
\end{eqnarray*}
\noindent
Now we can compute the two--form $\Omega$
\begin{eqnarray*}
\frac{1}{2}\Omega _{mn}\gamma ^{mn} &=&\epsilon \left[ \left( \cos\theta\, 
C^{\left( 2\right) }-\frac{1}{2}\sin\theta\, 
\left[ \gamma ^{2}\gamma ^{3},C^{\left( 2\right) }\right] \right) \,\gamma
^{4}\cdots \gamma ^{9}\right] _{2\text{--form}}\, \\
&=&\epsilon \sin\theta\,\gamma^0\gamma^2 -\epsilon \cos\theta\,\gamma ^{0}\gamma ^{3}
\end{eqnarray*}
\noindent
so that, finally, 
\begin{equation*}
\Omega =\epsilon \,\sin \theta\, dx^{0}\wedge dx^{2}
-\epsilon \,\cos\theta\, dx^{0}\wedge dx^{3}.
\end{equation*}
\noindent
We are now in a position to compute the two sides of (\ref{fullSduality}).
The LHS reads 
\begin{equation*}
\delta F\left( \sin \theta \,dx^{0}dx^{1},\epsilon \,dx^{1}dx^{2},0,0\right) =
\frac{1}{2}\epsilon \,dx^{1}dx^{2}\text{\thinspace }.
\end{equation*}
\noindent
To compute the RHS, it is convenient to first compute 
\begin{equation*}
F+\delta F\left( f\,dx^{2}dx^{3},0,\epsilon \,dx^{1}dx^{2},0\right) =\tan
\theta\  dx^{2}dx^{3}-\frac{\epsilon }{2\cos\theta}
dx^{0}dx^{3}\, ,
\end{equation*}
\noindent
which is mapped, under $S$, to 
\begin{equation*}
\sin\theta\, dx^{0}dx^{1}+\frac{\epsilon }{2}dx^{1}dx^{2}+\mathcal{O}\left(
\epsilon ^{2}\right) .
\end{equation*}
\noindent
Therefore, the RHS of (\ref{fullSduality}) is also equal to $\frac{1}{2}
\epsilon \,dx^{1}dx^{2}$, thus proving, in this example, $S$--duality to all
orders in $F$. For this check, it is crucial that the $C$ field induces a 
boundary deformation $\delta F$.

Let us consider a second example, where we set $C^{\left( 2\right) }=0$
and we consider a background with a constant $C^{\left( 4\right) }$ field,
\begin{equation*}
C^{\left( 4\right) }=\epsilon \,dx^{0}\wedge \cdots \wedge dx^{3}\, ,
\end{equation*}
and the same $U\left( 1\right) $ field strength $F=f\,dx^{2}\wedge dx^{3}$.
First of all we must compute the two--form $\Omega $ given this time by 
\begin{equation*}
\frac{1}{2}\,\Omega _{mn}\gamma ^{mn}=
-\epsilon \,\sin\theta\,\,\gamma ^{2}\gamma ^{3}\gamma
^{0}\cdots \gamma ^{9}\ ,
\end{equation*}
or by 
\begin{equation*}
\Omega =\epsilon \,\sin\theta \,dx^{2}\wedge dx^{3}\,.
\end{equation*}
Therefore, using equation (\ref{deltaF}), we conclude that 
\begin{equation}
\delta F\left( f\,dx^{2}dx^{3},0,0,\epsilon \,dx^{0}\cdots dx^{3}\right) =
\frac{\epsilon }{2}\,\frac{\tan\theta}{\cos\theta}\,
dx^{2}\wedge dx^{3}\,.  
\label{nextdf}
\end{equation}
It is quite easy to compute the $S$--dual field strength $S\left(
F+\delta F\right) $ using formulae (\ref{s1}) and (\ref{s2}), with $f$ replaced by 
$f\left( 1+\frac{1}{2}\epsilon \cos ^{-1}\theta\right ) $ 
\begin{eqnarray*}
S\left( F+\delta F\right)  &=&\sin \left( \arctan \left( \tan\theta\, 
+\frac{\epsilon }{2}\,\frac{\tan\theta}{\cos\theta}\right)\right) dx^{0}\wedge dx^{1} \\
&\simeq &\sin\theta \, dx^{0}\wedge dx^{1}+\frac{\epsilon }{2}
\sin\theta\,\cos\theta\, dx^{0}\wedge dx^{1}\,,
\end{eqnarray*}
so that the RHS of equation (\ref{fullSduality}) is given by 
\begin{equation}
\frac{\epsilon }{2}\sin\theta\, \cos\theta\,
dx^{0}\wedge dx^{1}\,.  \label{lastRHS}
\end{equation}
The LHS of (\ref{fullSduality}) reads, on the other hand, 
\begin{equation}
\delta F\left( \sin\theta\, dx^{0}dx^{1},0,0,\epsilon
\,dx^{0}\cdots dx^{3}\right) .  \label{finaldf}
\end{equation}
This is easily computed by noting that all the formulae which are valid for
the Euclidean directions of the brane world--volume are also valid for the
Minkowski directions, with the replacement of all the trigonometric
functions with the corresponding hyperbolic functions. Therefore, if we
define the angle $\widetilde{\theta }$ by 
\begin{equation*}
\tanh \widetilde{\theta }=\sin \theta \,,
\end{equation*}
then the result for (\ref{finaldf}) is given by the (hyperbolic version of)
equation (\ref{nextdf}) 
\begin{equation*}
\frac{\epsilon }{2}\,\frac{\tanh \widetilde{\theta }}
{\cosh \widetilde{\theta}}\,dx^{0}\wedge dx^{1}\,.
\end{equation*}
Noting that 
\begin{equation*}
\frac{1}{\cosh \widetilde{\theta }}=\cos \theta 
\end{equation*}
we recover the RHS (\ref{lastRHS}), as we wanted to show.

\section{Conclusions and Future Directions}

We have shown in this paper that, contrary to naive expectations, closed 
string backgrounds with constant NSNS or RR potentials do not satisfy, in 
the presence of $D$--branes, the open/closed equations of motion,
since $B$ and $C$--fields induce non--vanishing identical open string tadpoles
and currents. Open strings then must react also in the presence of gauge 
trivial RR fields\footnote{The fact that the world--sheet string action could 
include couplings to the RR gauge potentials, was previously observed in the 
different context of matrix string theory in weakly curved backgrounds 
\cite{Schiappa, BJL}.}. Careful computations are carried out both in the RNS 
and in the pure spinor covariant formalisms in order to check these statements. 
While the infinitesimal results are quite simple, the full non--linear 
deformation produced by the closed string background NSNS and RR fields is 
obtained by solving an algebraically complex differential equation, which 
requires a perturbative treatment. We have analyzed this equation for the 
case of a $D3$--brane and we have shown that our result is indeed compatible 
with $S$--duality to all orders in the background fields. This is a strong 
check of the validity of the method.

A schematic summary of our reasoning is the following. One starts with a 
given $D$--brane with boundary conditions ${\mathcal{B}}$. After turning on 
a closed string field background, $\Psi$, one is thus led to different 
boundary conditions ${\mathcal{B}}'$. When looking at this situation 
from the $S$--dual point of view, we have the following. One begins 
with some $S$--dual boundary conditions $\tilde{\mathcal{B}}$, which 
are driven to new boundary conditions $\tilde{\mathcal{B}'}$ after 
turning on the $S$--dual closed string background $\tilde{\Psi}$. Our 
main point is that the boundary conditions $\tilde{\mathcal{B}'}$ 
should be $S$--dual to ${\mathcal{B}}'$.

Even though the basic results can be found in both the RNS and pure spinor
formalisms, the latter covariant formalism is much simpler and more
powerful. Let us briefly comment on this matter. The pure spinor formalism and
the RNS formalism are related via a specific map \cite{Berkovits-4}, where
the RNS operator
\begin{equation*}
c\partial c\partial ^{2}c\ e^{-2\phi }
\end{equation*}
\noindent 
is mapped to the pure spinor operator
\begin{equation*}
\lambda \gamma_{m} \theta\ \lambda \gamma_{n} \theta\ \lambda \gamma_{p}
\theta\ \theta \gamma^{mnp} \theta\ , 
\end{equation*}
\noindent 
yielding the unusual zero mode saturation of the pure spinor
formalism \cite{Berkovits-1}. As to the BRST operator, the pure spinor
\begin{equation*}
Q=\oint \lambda ^{\alpha }d_{\alpha }
\end{equation*}
\noindent 
is mapped, under the field redefinition of \cite{Berkovits-4}, to
\begin{equation*}
Q^{\prime}= Q_{\mathrm{RNS}} + \oint \eta\ . 
\end{equation*}
\noindent 
Here $Q_{\mathrm{RNS}}$ is the RNS BRST operator (\ref
{QBRSTRNS}), and one observes that the cohomology of $Q^{\prime }$ in the 
\textit{large} Hilbert space exactly coincides with the cohomology of $Q_{%
\mathrm{RNS}}$, which acts in the \textit{small} Hilbert space \cite
{Berkovits-4}.

In the large Hilbert space, where $Q^{\prime }$ acts, picture changing is a
gauge transformation. So, although physical states can be represented by
vertex operators in different pictures in the cohomology of $Q_{\mathrm{RNS}}
$, all such vertex operators are equivalent in the cohomology of $Q^{\prime }
$ (as needs to be, due to the well known fact that space--time supersymmetry
in the RNS formalism only closes modulo picture changing) \cite{Berkovits-4}. 
In light of this result, one observes that the pure spinor formalism is
summing over all the different possible pictures of the RNS formalism. That
is why this calculation seems to arise in a more natural way from the pure
spinor formalism rather than the RNS formalism.

Let us comment on the peculiar factor of $1/2$ which differs from the
usual lore, and which is crucial in our test of $S$--duality. That 
such a factor needs to be present is clear from a technical point of 
view. While in the bosonic string a constant $B$ solves the equations 
of motion, in the superstring there is a current so that this will no 
longer be a solution to the equations of motion (the reason for this 
is, of course, the fact that the world--sheet theories are different). 
As we have seen this is true for canonically normalized fields, as the 
combination $F+B$ is still gauge invariant at the $\sigma$--model level.
Regarding the reason why 
this factor has never been encountered before, we just note in here that 
most of the literature on noncommutative gauge theory only deals with the 
boundary CFT, and what is called $B$ parametrizes the boundary condition, 
just like $F$ does in section $6$. It remains an open problem to find a 
clearer physical manifestation of this factor.

A point that is important to study further is the relation with the usual
supergravity solutions which represent $D$--branes (and in particular 
$D3$--branes) immersed in a constant $B$--field \cite{BMM,CostaPapa}, together
with their decoupling limit which should be dual to noncommutative SYM on
the brane \cite{Malda-Russo}. To be able to compare the two results, 
one should thus extend our results beyond the small $B$--field regime, 
in order to match it to the SUGRA solutions which are most relevant in a 
decoupling regime, with a large $B$ field. It seems, as well, that in 
this supergravity setting one will also not be able to distinguish 
the factor of $1/2$: by performing a gauge transformation on the 
supergravity solution one can eliminate completely the fields at 
infinity. The only invariant object is still the boundary condition 
and we are therefore in a situation similar to the one in the 
preceding paragraph.

Finally one should recall that the usual gauge invariance of Born--Infeld is
compatible, in the presence of branes wrapped on tori, with the periodicity
of $B$ under $T$--duality and the quantization of $F$ due to the fact that
the gauge group is compact. One should therefore further analyze our methods
in the presence of compact tori, and understand the behavior of our results
under $T$--duality.

There is still much work to be done in order to fully understand open string
physics in the presence of RR fields. One thing that would be of interest
would be to write down the open string sigma model describing our situation,
and actually solving it. That would be the ``integrated vertex operator''
version of the results in our paper. Another point of interest would be to
completely solve the differential equation which deals with the non--linear
open string deformation. This would determine the open string parameters $G$
and $\Theta$ for arbitrary NSNS and RR closed string backgrounds (pure
gauge). Finally, one should attempt to attack this problem for varying RR fields, 
along the lines of the work in \cite{Ho-Yeh, Cornalba-Schiappa-2} for the
$B$--field case. In this regard, the solution to the source equation 
in appendix B could serve as a starting ground in order to generalize 
\cite{Cornalba-Schiappa-2} to this superstring setting. If this line of 
reasoning can be pushed far enough, one could envisage translating the physics 
of the massless modes of open strings in arbitrary NSNS and RR backgrounds to 
noncommutative gauge theory with a very specific (associative or not) star 
product deformation.

\section*{Acknowledgements}

We would like to thank Nathan Berkovits for collaboration at the initial 
stages of this research and for useful discussions and correspondence. 
We would also like to thank Rodolfo Russo and Ashoke Sen for comments.
LC has been supported by the Stichting voor Fundamenteel Onderzoek der Materie
and by a Marie Curie Fellowship under the European 
Commission's Improving Human Potential programme (HPMF-CT-2002-02016),
and would like to thank the Isaac Newton Institute for Mathematical Sciences 
in Cambridge for hospitality during a vital stage of this project. MSC has 
been partially supported by a Marie Curie Fellowship under the European 
Commission's Improving Human Potential programme (HPMFCT--2000--00508), 
and also wishes to thank the Isaac Newton Institute for Mathematical Sciences 
in Cambridge for hospitality. RS is supported in part by funds provided by the 
Funda\c c\~ao para a Ci\^encia e a Tecnologia, under the grant SFRH/BPD/7190/2001 
(Portugal) and would like to thank the Caltech--USC Center for Theoretical 
Physics for a very nice hospitality during the course of this work.

\vfill

\eject

\appendix

\section{Spinor Conventions}

Let $\Gamma ^{m}$ be the $10$ dimensional $32\times 32$ gamma matrices,
satisfying 
$$
\left\{ \Gamma ^{m},\Gamma ^{m}\right\} =2g^{mn}.
$$
\noindent
They can be constructed starting from the \textit{real, symmetric, 
Euclidean} $9$ dimensional $16\times 16$ matrices $\gamma ^{i}$ (with $i=1,
\ldots,9$), chosen so that $\gamma^{1} \cdots \gamma^{9}=1$. Using the fact 
that $C\left( 9,0\right) \times C\left( 1,1\right) =C\left( 10,1\right)$, 
one writes
\begin{eqnarray*}
\Gamma ^{0} &=&1\otimes i\sigma _{y}=\left( 
\begin{array}{cc}
0 & 1 \\
-1 & 0
\end{array}
\right) , \\
\Gamma ^{i} &=& \gamma ^{i}\otimes \sigma _{x}=\left( 
\begin{array}{cc}
0 & \gamma ^{i} \\ 
\gamma ^{i} & 0
\end{array}
\right) , \\
\overline{\Gamma } &=&\Gamma ^{0}\cdots \Gamma ^{9}=1\otimes \sigma
_{z}=\left( 
\begin{array}{cc}
1 & 0 \\ 
0 & -1
\end{array}
\right) .
\end{eqnarray*}
\noindent
In particular we can define the following matrices
$$
\left( \gamma ^{m}\right) ^{\alpha \beta } =1,\gamma _{i}\, , \ \ \ \ \ 
\ \ \ \ \ \ \ \ \ \ \ \ \ 
\left( \gamma ^{m}\right) _{\alpha \beta } =-1,\gamma _{i}\, ,
$$
\noindent
so that 
$$
\Gamma ^{m}=\left( 
\begin{array}{cc}
0 & \left( \gamma ^{m}\right) ^{\alpha \beta } \\ 
\left( \gamma ^{m}\right) _{\alpha \beta } & 0
\end{array}
\right) .
$$
\noindent
Chiral and anti--chiral spinors will then carry a greek index
$$
\psi^\alpha\, , \ \ \ \ \ \ \ \ \ \ \ \ \ \ \ \ \ \ \psi_\alpha\, ,
$$
\noindent
and indices can only be raised and lowered with the matrices $\gamma^m$. 
Finally the basic self--duality relations read
\begin{eqnarray*}
&&(\gamma^0 \cdots \gamma^9)^\alpha{}_\beta = \delta^\alpha{}_\beta\, , \\
&&(\gamma^0 \cdots \gamma^9)_\alpha{}^\beta = 
-\delta_\alpha{}^\beta\, .
\end{eqnarray*}

\section{Solution of the Source Equation $Q\Phi_c=-J$ in the Covariant 
Formalism}

The basic equation which we want to solve is
\begin{equation}\label{a0}
Q (\la A) = (\la D) (\la A) = J ,
\end{equation}
\noindent
or, with explicit indices,
$$
\la^\a \la^\b D_\a A_\b = \frac{1}{2} \la^\a \la^\b J_{\a\b},
$$
\noindent
where $\la$ satisfies the pure spinor constraint $\la \g^m \la = 0$. The 
current $J$ is BRST closed ---\textit{i.e.}, it satisfies
\begin{equation}\label{a1}
QJ = \frac{1}{2} \la^\a \la^\b \la^\g  D_\a J_{\b\g} = 0.
\end{equation}

\subsection{Properties of the Current $J$}

First, let us analyze the basic consistency properties of the current $J$. 
Since $\la$ is a pure spinor, the current consists only of a self--dual 
five--form part
\beas
J_{\a\b} &=& \frac{1}{5!}\,J_{m_1\cdots m_5}\,
\left(\g^{m_1\cdots m_5}\right)_{\a\b}\, ,
\\
J_{m_1\cdots m_5} &=& \frac{1}{16}\,J_{\a\b} 
\left(\g_{m_1\cdots m_5}\right)^{\a\b}\, .
\eeas
\noindent
Now consider the derivative $D_\a J_{\b\g}$. The totally symmetrized part 
$D_{(\a} J_{\b\g)}$ is constrained by (\ref{a1}) to be
\bea\label{a2}
\frac{1}{4} \left( D_{\a} J_{\b\g} + {\mathrm{cyclic}}_{\a\b\g} \right) =
\g_{\a\b}^m J_{m\g} + {\mathrm{cyclic}}_{\a\b\g}\, ,
\eea
\noindent
where we have defined
\beas
J_{m\g} = \frac{1}{40} \g_m^{\a\b} D_\a J_{\b\g}\, ,
\ \ \ \ \ \ \ \ \ \ \ \ \ \ 
J_{\g m} = - J_{m\g}\, .
\eeas
\noindent
The structure of equation (\ref{a2}) is fixed by group theoretic arguments. 
The only thing one needs to check is the relative normalization of the two 
sides of the equation. To do this, we contract both sides with $\g_p^{\a\b}$. 
The left--hand side becomes $20 J_{p\g}$. The right--hand side requires more 
work. First we wish to show that
$$
J_{m\g} = \frac{1}{480} \left(D\g^{n_1\cdots n_4}\right)_\g\, 
J_{mn_1\cdots n_4}\, .
$$
\noindent
This follows from the following computation
\beas
\g_m^{\a\b} D_\a J_{\b\g} &=& \frac{1}{5!} \left( D \g_m 
\g^{n_1\cdots n_5}\right)_\g\, J_{n_1\cdots n_5}\\
&=& \frac{1}{5!} D \left( \g_m{}^{n_1\cdots n_5}+ 5\eta_m{}^{n_1}
\g^{n_2\cdots n_5}  \right)_\g\, J_{n_1\cdots n_5}\\
&=& \frac{1}{12} \left( D \g^{n_2\cdots n_5}  \right)_\g\, 
J_{mn_2\cdots n_5}\, ,
\eeas
\noindent
where we have used the Hodge duality relation,
$$
J_{n_1\cdots n_5}\left(\g^{m n_1\cdots n_5}\right)^\b{}_\g = 
5 J^m{}_{n_2\cdots n_5}\left(\g^{n_2\cdots n_5}\right)^\b{}_\g\, .
$$
\noindent
Using this result we can evaluate the following quantity
\beas
\g_p^{\a\b} \g^m_{\b\g} J_{m\a} 
&=& \frac{1}{40\cdot 5!} \left( D\g_m\g^{n_1\cdots n_5} \g_p\g^m 
\right)_\g J_{n_1\cdots n_5}\\
&=& \frac{1}{40\cdot 5!} 
\left(D \g_m \left[ \g^{n_1\cdots n_5}{}_p + 5\g^{n_1\cdots n_4} 
\eta^{n_5}{}_p\right]\g^m\right)_\g J_{n_1\cdots n_5}\\
&=& \frac{1}{40\cdot 5!} D \left( 2 \g_p{}^{n_1\cdots n_5 } + 10 
\g^{n_2\cdots n_5 } \eta^{n_1}{}_p\right)_\g J_{n_1\cdots n_5}\\
&=& 2 J_{p\g}\, ,
\eeas
\noindent
which implies that the contraction of the right--hand side of (\ref{a2}) 
with $\g_p^{\a\b}$ is also $20 J_{p\g}$.

For future purposes we define
$$
J_{mn} = \frac{1}{32} \g_m^{\a\b} D_\a J_{\b n}\, .
$$
\noindent
We claim that 
$$
J_{mn} = -J_{nm} = \frac{1}{3840} \left( D\g^{p_1p_2p_3}D \right) 
J_{mnp_1p_2p_3}\, .
$$
\noindent
It is clear from the definition that
$$
J_{mn} = -\frac{1}{5!\cdot 1280} \left( D\g_m \g^{p_1\cdots p_5} 
\g_n D\right) J_{p_1\cdots p_5}\, .
$$
\noindent
Moreover, using the fact that $D_\a D_\b$ is a bi--spinor which does not 
have the $5$--form part, we conclude that we can use
\beas
\g_m\g^{p_1\cdots p_5} \g_n J_{p_1\cdots p_5} &=& \left( 
\g_m{}^{p_1\cdots p_5}{}_n J_{p_1\cdots p_5}  - 20 \g^{p_1\cdots p_3}
J_{mn p_1 p_2 p_3} \right) + 5{\rm-form}\\
&=& -40 \g^{p_1\cdots p_3} J_{mn p_1 p_2 p_3} + 5{\rm-form}\, ,
\eeas
\noindent
thus confirming the claim.

\subsection{First Bianchi Identity and General Solution for $A_\a$}

Now, let us start to analyze the constraints coming from the Bianchi 
identities. First of all we define the field strengths as
\beas
F_{\a\b} &=& \left\{ \nabla_\a , \nabla_\b \right\} - 4\g_{\a\b}^m 
\nabla_m\, , \\
F_{m\a} &=& -F_{\a m} = \left[ \nabla_m ,\nabla_\a \right]\, , \\
F_{mn} &=& \left[ \nabla_m ,\nabla_n \right]\, ,
\eeas
\noindent
where
$$
\nabla_\a = D_\a + A_\a\, , \ \ \ \ \ \ \ \ \ \ \ \ \ \ \ 
\nabla_m = \partial_m + A_m\, .
$$
\noindent
The basic equation (\ref{a0}) then reads
$$
F_{\a\b} = J_{\a\b}\, .
$$
\noindent
We consider the first Bianchi identity
\begin{equation*}
\left[ \left\{ \nabla _{\alpha },\nabla _{\beta }\right\} ,\nabla _{\gamma }
\right] +\left[ \left\{ \nabla _{\beta },\nabla _{\gamma }\right\} ,\nabla
_{\alpha }\right] +\left[ \left\{ \nabla _{\beta },\nabla _{\gamma }\right\}
,\nabla _{\alpha }\right] = 0 ,
\end{equation*}
\noindent
which implies that
$$
\gamma _{\alpha \beta }^{m}F_{m\gamma }+{\mathrm{cyclic}}_{\a\b\g} = 
\frac{1}{4} \left( D_\a J_{\b\g} + {\mathrm{cyclic}}_{\a\b\g}\right) .
$$
\noindent
We therefore conclude that
\beas
F_{m\a} &=& J_{m\a} - 2 \left( \g_m \right)_{\a\b} W^\b\, , \\
W^\a &=& \frac{1}{20} \left( \g^m \right)^{\a\b}\left(F_{\b m} - 
J_{\b m}\right)\, .
\eeas
\noindent
We may also use the above equation to solve, in Lorentz gauge, for $A_\a$ as 
a function of $W$ and $J$. In fact, since
$$
\partial^m F_{m\a} = \square A_\a - D_\a \left( \partial\cdot A\right) = 
-2 \left( \delsl W\right)_\a + \partial^m J_{m\a}\, ,
$$
\noindent
and if
$$
\partial\cdot A = 0 ,
$$
\noindent
one obtains that
$$
A_\a = \frac{1}{\square} \left[  -2 \left( \delsl 
W\right)_\a + \partial^m J_{m\a} \right].
$$

\subsection{Second Bianchi Identity and the Computation of $D_\alpha W^\beta$}

Next we consider the second Bianchi identity
\begin{equation*}
\left[ \left\{ \nabla _{\alpha },\nabla _{\beta }\right\} ,\nabla _{m}\right]
+\left\{ \left[ \nabla _{m},\nabla _{\alpha }\right] ,\nabla _{\beta
}\right\} -\left\{ \left[ \nabla _{\beta },\nabla _{m}\right] ,\nabla
_{\alpha }\right\} =0,
\end{equation*}
\noindent
which reads
\bea\label{a5}
4\gamma _{\alpha \beta }^{n}F_{nm}&=& 2 \left[ \left( \gamma_{m} 
\right) _{\alpha \gamma}D_{\beta }W^{\gamma }+\left( \gamma_{m} 
\right)_{\beta \gamma }D_{\alpha}W^{\gamma }\right] +
\\ &&+D_\a J_{\b m} + D_\b J_{\a m} + \partial_m J_{\a\b}\, .\notag
\eea
\noindent
Contracting this equation with $\g^{\a\b}_p$ we get
\beas
DW &=& 0\, ,\\
F_{p m} &=& J_{p m} + \frac{1}{16} D\g_{p m} W \, .
\eeas
\noindent
Now we look at the five--form part of equation (\ref{a5}) by contracting 
with $(\g^{n_1\cdots n_4m})^{\a\b}$. This gives
$$
24D\g^{n_1\cdots n_4} W =- 16 \partial_m J^{mn_1\cdots n_4}+
\frac{1}{240} \left( D\g^{n_1\cdots n_4m} \g^{p_1\cdots p_4} D\right) 
J_{mp_1\cdots p_4}\, .
$$
\noindent
To simplify the expression $D\g^{n_1\cdots n_4m} \g^{p_1\cdots p_4} D$ we use 
the fact that $D_\alpha D_\beta$ has only a one--form and a three--form part, 
together with the self--duality of the five--form $J_{m_1\cdots m_5}$. More 
precisely one can check that
\begin{eqnarray*}
(D\g^{n_1\cdots n_4 m}\g^{p_1\cdots p_4} D) J_{mp_1\cdots p_4} &=& 
-4\cdot 4!\, (D\g_m D) J^{mn_1\cdots n_4}\\
&& +4\cdot 4!\, (D \g^{ij[n_1} D) J^{n_2n_3n_4]}{}_{ij}\, .
\end{eqnarray*}
\noindent
Using the fact that
$$
D\g_m D = 32\,\partial_m\, ,
$$
\noindent
we conclude that
$$
D\g^{n_1\cdots n_4} W =- \frac{6}{5} \partial_m J^{mn_1\cdots n_4}+
\frac{1}{60} \left(D\gamma^{mp[n_1}D\right) J^{n_2n_3n_4]}{}_{mp}\, .
$$

\subsection{Third Bianchi Identity and Expression for $\delsl W$}

To calculate $\delsl W$ we use the expansion
$$
D_\n W^\beta = \frac{1}{16\cdot 2!} (\gamma_{mn})_\n{}^\beta ( D\gamma^{mn}W) 
+ \frac{1}{16\cdot 4!} (\gamma_{n_1\cdots n_4})_\n{}^\beta (
D\gamma^{n_1\cdots n_4}W)
$$
\noindent
to compute
\begin{eqnarray*}
16 (\delsl W)_\alpha &=& \frac{1}{2}(\gamma^p)_{\a\b} D_\m D_\n W^\b 
(\g_p)^{\m\n}\\
&=& \frac{3}{32} (D\gamma_{mn} )_\alpha ( D\gamma^{mn}W) + \frac{1}{384} 
(D\gamma_{n_1\cdots n_4} )_\alpha ( D\gamma^{n_1\cdots n_4}W)\\
&=& \frac{3}{2} (D\g_{mn})_\a F^{mn}-\frac{3}{2} (D\g_{mn})_\a J^{mn}
 -\frac{1}{320} \partial_m (D\g_{n_1\cdots n_4})_\a J^{mn_1\cdots n_4} \\
&&+\frac{1}{23040} (D\gamma_{n_1\cdots n_4})_\a (D\gamma^{mpn_1}D)
J^{n_1n_2n_3}{}_{mp}\, ,
\end{eqnarray*}
\noindent
where we have used the fact that $\gamma_p \gamma^{n_1\cdots n_p} \gamma^p = 
(-)^p (10-2p)\gamma^{n_1\cdots n_p}$. We will also need to consider the 
consequences of the Bianchi identity
$$
[\nabla_\a, [\nabla_m,\nabla_n]] + [\nabla_m, [\nabla_n,\nabla_\a]] +
[\nabla_n, [\nabla_\a,\nabla_m]] = 0,
$$
\noindent
which implies
$$
-D_\a F_{mn} = 2\,\, {}_\a(\g_m \partial_n - \g_n \partial_m) W +
\partial_m J_{n\a} - \partial_n J_{m\a}\, .
$$
\noindent
Using the above expression, we may then compute
\begin{eqnarray*}
(D\g_{mn})_\a F^{mn} &=& -2 (\g^{mn})^\b{}_\a \partial_m J_{n\b} - 
4(\partial_n W\g_m\g^{mn})_\a \\
&=& -2 (\g^{mn})^\b{}_\a \partial_m J_{n\b} - 36 (\delsl W)_\a\, ,
\end{eqnarray*}
\noindent
where we have used that $\g_m \gamma^{mn} = 9 \g^n$. We then arrive at 
the result for $\delsl W$ given by
\begin{eqnarray*}
70 (\delsl W)_\alpha &=& -3 (\g^{mn})^\b{}_\a \partial_m 
J_{n\b} -\frac{3}{2} (D\g_{mn})_\a J^{mn}
-\frac{1}{320} \partial_m (D\g_{n_1\cdots n_4})_\a J^{mn_1\cdots n_4} \\
&&+\frac{1}{23040} (D\gamma_{n_1\cdots n_4})_\a (D\gamma^{mpn_1}D)
J^{n_2n_3n_4}{}_{mp}\, .
\end{eqnarray*}
\noindent
In order to simplify the expression given above, we note that
\begin{eqnarray*}
\partial_m J_{n\b} (\g^{mn})^\b{}_\a &=& \frac{1}{480} 
(D\g^{n_1\cdots n_4}\g^{mn})_\a \partial_m J_{nn_1\cdots n_4}\\
&=& \frac{1}{480} \partial_m (D\g_{n_1\cdots n_4})_\a J^{mn_1\cdots n_4} \\
&=& \partial^m J_{m\alpha}\, ,
\end{eqnarray*}
\noindent
so that we arrive at the final expression
\begin{eqnarray*}
70 (\delsl W)_\alpha &=& -\frac{9}{2} \partial^m J_{m\a}
-\frac{3}{2} (D\g_{mn})_\a J^{mn}\\
&&+\frac{1}{23040} (D\gamma_{n_1\cdots n_4})_\a (D\gamma^{mpn_1}D)
J^{n_2n_3n_4}{}_{mp}\, .
\end{eqnarray*}
\noindent
We therefore conclude that the expression for $\square A_\alpha$ is given 
by
\begin{eqnarray*}
70 \square A_\alpha &=& 79 \partial^m J_{m\a}+ 3 (D\g_{mn})_\a J^{mn}\\
&&-\frac{1}{11520} (D\gamma_{n_1\cdots n_4})_\a (D\gamma^{mpn_1}D)
J^{n_2n_3n_4}{}_{mp}\, .
\end{eqnarray*}


\vfill

\eject

\bibliographystyle{plain}

\begin{thebibliography}{10}

\bibitem{FMS}
D.~Friedan, E.~Martinec and S.~Shenker,
\textit{Conformal Invariance, Supersymmetry and String Theory}, Nucl.\
Phys.\ \textbf{B271} (1986) 93.

\bibitem{CFQS}
J.~Cohn, D.~Friedan, Z.~Qiu and S.~Shenker,
\textit{Covariant Quantization of Supersymmetric String Theories: The
Spinor Field of the Ramond--Neveu--Schwarz Model}, Nucl.\ Phys.\
\textbf{B278} (1986) 577.

\bibitem{KLLSW}
V.~A.~Kosteleck\'y, O.~Lechtenfeld, W.~Lerche, S.~Samuel and S.~Watamura, 
\textit{Conformal Techniques, Bosonization and Tree--Level String
Amplitudes}, Nucl.\ Phys.\ \textbf{B288} (1987) 173.

\bibitem{KLS}
V.~A.~Kosteleck\'y, O.~Lechtenfeld and S.~Samuel,
\textit{Covariant String Amplitudes on Exotic Topologies to One Loop},
Nucl.\ Phys.\ \textbf{B298} (1988) 133.

\bibitem{Klebanov-Thorlacius}
I.~R.~Klebanov and L.~Thorlacius,
\textit{The Size of $p$--Branes}, \texttt{hep-th/9510200}.

\bibitem{GHKM}
S.~S.~Gubser, A.~Hashimoto, I.~R.~Klebanov and J.~M.~Maldacena, 
\textit{Gravitational Lensing by $p$--Branes}, Nucl.\ Phys.\ \textbf{B472}
(1996) 231, \texttt{hep-th/9601057}.

\bibitem{Garousi-Myers}
M.~R.~Garousi and R.~C.~Myers, 
\textit{Superstring Scattering from $D$--Branes}, Nucl.\ Phys.\
\textbf{B475} (1996) 193, \texttt{hep-th/9603194}.

\bibitem{Hashimoto-Klebanov}
A.~Hashimoto and I.~R.~Klebanov,
\textit{Decay of Excited $D$--Branes}, Phys.\ Lett.\ \textbf{B381} (1996)
437, \texttt{hep-th/9604065}.

\bibitem{BIANCHI}
M.~Bianchi, G.~Pradisi and A.~Sagnotti,
\textit{Toroidal Compactification and Symmetry Breaking in Open
String Theories},
Nucl.\ Phys.\ \textbf{B376} 365 (1992).

\bibitem{VFPSLR}
P.~Di~Vecchia, M.~Frau, I.~Pesando, S.~Sciuto, A.~Lerda and R.~Russo, 
\textit{Classical $p$--Branes from Boundary State}, Nucl.\ Phys.\
\textbf{B507} (1997) 259, \texttt{hep-th/9707068}.

\bibitem{BVFLPRS}
M.~Bill\'o, P.~Di~Vecchia, M.~Frau, A.~Lerda, I.~Pesando, R.~Russo and
S.~Sciuto, 
\textit{Microscopic String Analysis of the $D0$--$D8$ Brane System and Dual
R--R States}, Nucl.\ Phys.\ \textbf{B526} (1998) 199,
\texttt{hep-th/9802088}.

\bibitem{Polyakov}
D.~Polyakov, 
\textit{BRST Properties of New Superstring States}, \texttt{hep-th/0111227}.

\bibitem{Siegel}
W.~Siegel, 
\textit{Classical Superstring Mechanics}, Nucl.\ Phys.\ \textbf{B263}
(1985) 93.

\bibitem{Berkovits-1}
N.~Berkovits,
\textit{Super--Poincar\'e Covariant Quantization of the Superstring}, JHEP\
\textbf{0004} (2000) 018, \texttt{hep-th/0001035}.

\bibitem{Berkovits-Vallilo}
N.~Berkovits and B.~C.~Vallilo,
\textit{Consistency of Super--Poincar\'e Covariant Superstring Tree
Amplitudes}, JHEP\ \textbf{0007} (2000) 015, \texttt{hep-th/0004171}. 

\bibitem{Berkovits-2}
N.~Berkovits,
\textit{Cohomology in the Pure Spinor Formalism for the Superstring}, JHEP\
\textbf{0009} (2000) 046, \texttt{hep-th/0006003}.

\bibitem{Berkovits-3}
N.~Berkovits,
\textit{Covariant Quantization of the Superstring}, Int.\ J.\ Mod.\ Phys.\
\textbf{A16} (2001) 801, \texttt{hep-th/0008145}.

\bibitem{Berkovits-Chandia-1}
N.~Berkovits and O.~Chand\'\i a,
\textit{Superstring Vertex Operators in an $AdS_{5} \times S^{5}$
Background}, Nucl.\ Phys.\ \textbf{B596} (2001) 185,
\texttt{hep-th/0009168}.

\bibitem{Berkovits-4}
N.~Berkovits,
\textit{Relating the RNS and Pure Spinor Formalisms for the Superstring},
JHEP\ \textbf{0108} (2001) 026, \texttt{hep-th/0104247}.

\bibitem{Berkovits-Chandia-2}
N.~Berkovits and O.~Chand\'\i a,
\textit{Lorentz Invariance of the Pure Spinor BRST Cohomology for the
Superstring}, Phys.\ Lett.\ \textbf{B514} (2001) 394,
\texttt{hep-th/0105149}.

\bibitem{Berkovits-Howe}
N.~Berkovits and P.~Howe,
\textit{Ten--Dimensional Supergravity Constraints from the Pure Spinor
Formalism for the Superstring}, \texttt{hep-th/0112160}.

\bibitem{Berkovits-5}
N.~Berkovits,
\textit{Conformal Field Theory for the Superstring in a Ramond--Ramond
Plane Wave Background}, \texttt{hep-th/0203248}.

\bibitem{Berkovits-Chandia-3}
N.~Berkovits and O.~Chand\'\i a,
\textit{Massive Superstring Vertex Operator in $d=10$ Superspace},
\texttt{hep-th/0204121}.

\bibitem{Berkovits-Pershin}
N.~Berkovits and V.~Pershin,
\textit{Supersymmetric Born--Infeld from the Pure Spinor Formalism of 
the Open Superstring},
\texttt{hep-th/0205154}.

\bibitem{Berkovits-6}
N.~Berkovits,
\textit{ICTP Lectures on Covariant Quantization of the Superstring}, 
\texttt{hep-th/0209059}.

\bibitem{GPN1}
P.~A.~Grassi, G.~Policastro, M.~Porrati and P.~van~Nieuwenhuizen,
\textit{Covariant Quantization of Superstrings without Pure Spinor 
Constraints},
\texttt{hep-th/0112162}.

\bibitem{GPN2}
P.~A.~Grassi, G.~Policastro and P.~van~Nieuwenhuizen,
\textit{The Massless Spectrum of Covariant Superstrings},
\texttt{hep-th/0202123}.

\bibitem{GPN3}
P.~A.~Grassi, G.~Policastro and P.~van~Nieuwenhuizen,
\textit{On the BRST Cohomology of Superstring with/without Pure 
Spinors},
\texttt{hep-th/0206216}.

\bibitem{GPN4}
P.~A.~Grassi, G.~Policastro and P.~van~Nieuwenhuizen,
\textit{The Covariant Quantum Superstring and Superparticle from 
their Classical Actions},
\texttt{hep-th/0209026}.

\bibitem{ACNY}
A.~Abouelsaood, C.~G.~Callan, C.~R.~Nappi and S.~A.~Yost, 
\textit{Open Strings in Background Gauge Fields}, Nucl.\ Phys.\
\textbf{B280} (1987) 599.

\bibitem{CDS}
A.~Connes, M.~R.~Douglas and A.~Schwarz, 
\textit{Noncommutative Geometry and Matrix Theory: Compactification on
Tori}, JHEP\ \textbf{9802} (1998) 003, \texttt{hep-th/9711162}.

\bibitem{Cheung-Krogh}
Y-K.~E.~Cheung and M.~Krogh, 
\textit{Noncommutative Geometry from $D0$--Branes in a Background
$B$--Field}, Nucl.\ Phys.\ \textbf{B528} (1998) 185,
\texttt{hep-th/9803031}.

\bibitem{Chu-Ho}
C-S.~Chu and P-M.~Ho, 
\textit{Noncommutative Open String and $D$--Brane}, \texttt{hep-th/9812219}.

\bibitem{Schomerus}
V.~Schomerus, 
\textit{$D$--Branes and Deformation Quantization}, JHEP\ \textbf{9906}
(1999) 030, \texttt{hep-th/9903205}.

\bibitem{Cornalba-Schiappa-1}
L.~Cornalba and R.~Schiappa,
\textit{ Matrix Theory Star Products from the Born--Infeld Action}, Adv.\
Theor.\ Math.\ Phys.\ \textbf{4} (2000) 249, \texttt{hep-th/9907211}.

\bibitem{Seiberg-Witten}
N.~Seiberg and E.~Witten, 
\textit{String Theory and Noncommutative Geometry}, JHEP\ \textbf{9909}
(1999) 032, \texttt{ hep-th/9908142}.

\bibitem{Cornalba-2}
L.~Cornalba, 
\textit{$D$--Brane Physics and Noncommutative Yang--Mills Theory}, Adv.\
Theor.\ Math.\ Phys.\ \textbf{4} (2000) 271, \texttt{hep-th/9909081}.

\bibitem{Chu-Zamora}
C-S.~Chu and F.~Zamora, 
\textit{Manifest Supersymmetry in Noncommutative Geometry}, JHEP\
\textbf{0002} (2000) 022, \texttt{hep-th/9912153}.

\bibitem{GRS}
O.~J.~Ganor, G.~Rajesh and S.~Sethi,
\textit{Duality and Noncommutative Gauge Theory}, Phys.\ Rev.\ 
\textbf{D62} (2000) 125008, \texttt{0005046}.

\bibitem{GMMS}
R.~Gopakumar, J.~M.~Maldacena, S.~Minwalla and A.~Strominger,
\textit{$S$--Duality and Noncommutative Gauge Theory}, JHEP\ 
\textbf{0006} (2000) 036, \texttt{0005048}.

\bibitem{NOS}
C.~N\'u\~nez, K.~Olsen and R.~Schiappa,
\textit{From Noncommutative Bosonization to $S$--Duality}, JHEP\ 
\textbf{0007} (2000) 030, \texttt{0005059}.

\bibitem{Zwiebach-1}
B.~Zwiebach, 
\textit{Interpolating String Field Theories},
Mod.\ Phys.\ Lett.\ \textbf{A7} (1992) 1079, \texttt{hep-th/9202015}.

\bibitem{Gaberdiel-Zwiebach-1}
M.~R.~Gaberdiel and B.~Zwiebach, 
\textit{Tensor Constructions of Open String Theories I: Foundations},
Nucl.\ Phys.\ \textbf{B505} (1997) 569, \texttt{hep-th/9705038}.

\bibitem{Zwiebach-2}
B.~Zwiebach, 
\textit{Oriented Open--Closed String Theory Revisited},
Annals\ Phys.\ \textbf{267} (1998) 193, \texttt{hep-th/9705241}.

\bibitem{Gaberdiel-Zwiebach-2}
M.~R.~Gaberdiel and B.~Zwiebach, 
\textit{Tensor Constructions of Open String Theories II: Vector Bundles, 
$D$--branes and Orientifold Groups},
Phys.\ Lett.\ \textbf{B410} (1997) 151, \texttt{hep-th/9707051}.

\bibitem{Berkovits-10}
N.~Berkovits,
\textit{Manifest Electromagnetic Duality in Closed Superstring Field Theory},
Phys.\ Lett.\ \textbf{B388} (1996) 743,
\texttt{hep-th/9607070}.

\bibitem{Berkovits-11}
N.~Berkovits,
\textit{Ramond--Ramond Central Charges in the Supersymmetry Algebra of the 
Superstring},
Phys.\ Rev.\ Lett.\ \textbf{79} 1813 (1997),
\texttt{hep-th/9706024}.

\bibitem{GibbRash}
G.W.~Gibbons and D.A.~Rasheed, 
\textit{$SL(2,\mathbb{R})$ Invariance of Non--Linear Electrodynamics Coupled 
to an Axion and a Dilaton},
Phys.\ Lett.\  \textbf{B365} 46 (1996),
\texttt{hep-th/9509141}.

\bibitem{Tsey}
A.A.~Tseytlin,
\textit{Self--Duality of Born--Infeld Action and Dirichlet 3--Brane 
of Type IIB Superstring Theory},
Nucl.\ Phys.\  \textbf{B469} 51 (1996),
\texttt{hep-th/9602064}.

\bibitem{GreenGutp}
M.B.~Green and M.~Gutperle,
\textit{Comments on Three--Branes},
Phys.\ Lett.\  \textbf{B377} 28 (1996),
\texttt{hep-th/9602077}.

\bibitem{Schiappa}
R.~Schiappa, 
\textit{Matrix Strings in Weakly Curved Background Fields}, Nucl.\ Phys.\
\textbf{B608} (2001) 3, \texttt{hep-th/0005145}.

\bibitem{BJL}
D.~Brecher, B.~Janssen and Y.~Lozano,
\textit{Dielectric Fundamental Strings in Matrix String Theory},
\texttt{hep-th/0112180}.

\bibitem{BMM}
J.C.~Breckenridge, G.~Michaud and R.C.~Myers,
\textit{More $D$--Brane Bound States},
Phys.\ Rev. \ \textbf{D55} 6438 (1997),
\texttt{hep-th/9611174}.

\bibitem{CostaPapa}
M.S.~Costa and G.~Papadopoulos,
\textit{Superstring Dualities and $p$--Brane Bound States},
Nucl.\ Phys.\  \textbf{B510} 217  (1998),
\texttt{hep-th/9612204}.

\bibitem{Malda-Russo}
J.~M.~Maldacena and J.~G.~Russo,
\textit{Large $N$ Limit of Noncommutative Gauge Theories}
JHEP {\bf 9909}, 025 (1999),
\texttt{hep-th/9908134}.

\bibitem{Ho-Yeh}
P-M.~Ho and Y-T.~Yeh, 
\textit{Noncommutative $D$--Brane in Nonconstant NS--NS $B$--Field
Background}, Phys.\ Rev.\ Lett.\ \textbf{85} (2000) 5523,
\texttt{hep-th/0005159}.

\bibitem{Cornalba-Schiappa-2}
L.~Cornalba and R.~Schiappa,
\textit{Nonassociative Star Product Deformations for $D$--Brane
Worldvolumes in Curved Backgrounds}, Commun.\ Math.\ Phys.\ \textbf{225}
(2002) 33, \texttt{hep-th/0101219}.

\end{thebibliography}

\end{document}